\documentclass{article}

    \usepackage[preprint]{neurips_2023}

\usepackage[utf8]{inputenc} 
\usepackage[T1]{fontenc}    
\usepackage{hyperref}       
\usepackage{url}            
\usepackage{booktabs}       
\usepackage{amsfonts}       
\usepackage{nicefrac}       
\usepackage{microtype}      
\usepackage{xcolor}         
\usepackage{graphicx}
\usepackage{amsmath}
\usepackage{enumitem}
\usepackage{multicol}
\usepackage{multirow}
\usepackage{wrapfig}

\title{\method: Cross-data Biosignal Learning in the Wild}

\author{Chaoqi Yang$^{1}$, ~M. Brandon Westover$^{2,3}$, ~Jimeng Sun$^{1}$ \\ $^1$University of Illinois Urbana-Champaign, $^2$Harvard Medical School \\ $^3$Beth Israel Deaconess Medical Center \\
\texttt{\{chaoqiy2\}@illinois.edu}
}

\usepackage{xspace}

\newcommand{\bfF}{\mathbf{F}}

\newcommand{\bfW}{\mathbf{W}}
\newcommand{\bfE}{\mathbf{E}}
\newcommand{\bfH}{\mathbf{H}}
\newcommand{\bfX}{\mathbf{X}}

\newcommand{\bfI}{\mathbf{I}}

\newcommand{\bfS}{\mathbf{S}}
\newcommand{\RR}{\mathbb{R}}
\newcommand{\bfZ}{\mathbf{Z}}

\newcommand{\calL}{\mathcal{L}}

\newcommand{\predictor}{\mathrm{predictor}}

\newcommand{\percentile}{\mathrm{percentile}}
\newcommand{\softmax}{\mathrm{softmax}}
\newcommand{\attention}{\mathrm{Attention}}
\newcommand{\crossentropy}{\mathrm{CrossEntropyLoss}}
\newcommand{\method}{\texttt{BIOT}\xspace}
\setlist[itemize]{leftmargin=*}

\begin{document}

\maketitle

\begin{abstract}
Biological signals, such as electroencephalograms (EEG), play a crucial role in numerous clinical applications, exhibiting diverse data formats and quality profiles. Current deep learning models for biosignals are typically specialized for specific datasets and clinical settings, limiting their broader applicability. Motivated by the success of large language models in text processing, we explore the development of foundational models that are trained from multiple data sources and can be fine-tuned on different downstream biosignal tasks.

To overcome the unique challenges associated with biosignals of various formats, such as mismatched channels, variable sample lengths, and prevalent missing values, we propose a Biosignal Transformer (\method). The proposed \method model can enable cross-data learning with mismatched channels, variable lengths, and missing values by tokenizing diverse biosignals into unified "biosignal sentences". Specifically, we tokenize each channel into fixed-length segments containing local signal features, flattening them to form consistent "sentences". Channel embeddings and {\em relative} position embeddings are added to preserve spatio-temporal features.

The \method model is versatile and applicable to various biosignal learning settings across different datasets, including joint pre-training for larger models. Comprehensive evaluations on EEG, electrocardiogram (ECG), and human activity sensory signals demonstrate that \method outperforms robust baselines in common settings and facilitates learning across multiple datasets with different formats. Use CHB-MIT seizure detection task as an example, our vanilla \method model shows 3\% improvement over baselines in balanced accuracy, and the pre-trained \method models (optimized from other data sources) can further bring up to 4\% improvements.
\end{abstract}
\vspace{-2mm}
\section{Introduction}
\vspace{-2mm}

Biosignals, such as EEG and ECG, are multi-channel time series recorded at high sampling rates (e.g., 256Hz) in various healthcare domains, including sleep medicine, neurological and cardiovascular disease detection, and activity monitoring.
Deep learning (DL) models have demonstrated impressive success in automating biosignal diagnosis across diverse applications~\citep{yang2021self}, encompassing sleep stage classification~\citep{biswal2018expert,yang2021self,phan2022automatic}, emotion analysis via EEG~\citep{zhang2020investigation,suhaimi2020eeg}, action and motor imagery recognition \citep{venkatachalam2020novel}, acute stress detection through electrodermal activity~\citep{greco2021acute}, EEG-based seizure epilepsy classification~\citep{yangmanydg,jing2023development}, and ECG-driven cardiac arrhythmia detection~\citep{isin2017cardiac,parvaneh2019cardiac}.

Various deep learning methods have been applied to biosignal analysis. Some works use 1D convolutional neural networks (CNN) on raw signals~\citep{jing2023development,nagabushanam2020artifact,dar2020cnn}, while others preprocess the data with short-time Fourier transform (STFT) and employ 2D CNN models on the resulting spectrogram~\citep{yang2022atd,kim2020study,cui2020eeg}. Researchers also segment the signal and use a CNN segment encoder with a downstream sequence model~\citep{zhang2019cnnlstm,biswal2018expert,jing2020development,almutairi2021detection}, such as Transformer or recurrent neural networks (RNN), to capture temporal dynamics. Other approaches involve ensemble learning, feature fusion from multiple encoders~\citep{li2022motor}, and multi-level transformers to encode spatial and temporal features across and within channels~\citep{lawhern2018eegnet,song2021transformer,liu2021transformers}.


These models \citep{jing2023development,yang2021self,biswal2018expert,kostas2021bendr,du2022eeg,zhang2022self} 
predominantly focus on biosignal samples with fixed formats for specific tasks, while real-world data may exhibit mismatched channels, variable lengths, and missing values.  In this paper, our objective is to devise a flexible training strategy that can handle diverse biosignal datasets with varying channels, lengths, and levels of missingness.
For example, is it possible to transfer knowledge from abnormal EEG detection (a binary classification task with 64 channels and a 5-second duration, recorded at 256Hz) to improve another EEG task, such as seizure type classification (a multi-class task with 16 channels and a 10-second duration at 200Hz)? In reality, such data mismatches often arise from varying devices, system errors, and recording limitations. Additionally, it is also important to explore the potential of utilizing different unlabeled data.

To apply existing deep learning models to such settings of different biosignals, significant data processing is required to align the formats across multiple datasets. 
This may involve truncating or padding signals for consistent lengths~\citep{zhang2022self}, and imputing missing channels or segments~\citep{bahador2021reconstruction}. 
Such practices, however, may introduce unnecessary noise and shift data distributions, leading to poor generalization performance. 
Developing a flexible and unified model that accommodates biosignals with diverse formats can be advantageous. 


In our paper, we develop the biosignal transformer (\method) model (summarized in Figure~\ref{fig:bst}), which, to the best of our knowledge, is the first biosignal encoding model that can handle biosignals of various formats. Our motivation stems from the vision transformer (ViT) \citep{dosovitskiy2020image} and the audio spectrogram transformer (AST) \citep{gong2021ast}. The ViT model splits the image into a "sentence" of patches for image representation. The AST model splits the audio spectrogram into "sentence" for 1D audio representation. These "sentence" structures combined with Transformer \citep{vaswani2017attention} can handle variable-sized inputs. 

Compared to images (RGB or gray), audios, or natural languages, biosignals are more complicated primarily as it has multiple channels. It is non-trivial to transform diverse biosignals of various formats into unified "sentence" structures. This paper proposes \method to solve the challenge by a novel {\bf biosignal tokenization module} that segments each channel separately into tokens and then flattens the tokens to form consistent biosignal "sentences" (illustrated in Figure~\ref{fig:sentence}). With the design, our \method can enable the knowledge transfer cross different data in the wild and allow joint (pre-)training on multiple biosignal data sources. Our contributions are listed below.

\begin{itemize}
    \item {\bf Biosignal transformer (\method).} This paper proposes a biosignal encoding model \method by tokenizing biosignals of various formats into unified ``sentences.''
    \item {\bf Knowledge transfer across different data.} Our \method can enable joint (pre-)training and knowledge transfer across different biosignal datasets in the wild, which inspires the research of large foundation models for biosignals.
    \item {\bf Strong empirical performance.} We evaluate our \method on several unsupervised and supervised EEG, ECG, and human activity sensory datasets. 
    Results show that \method outperforms baseline models and can utilize the models pre-trained on other data to benefit the current task.
\end{itemize}

\section{\method: Biosignal Transformer}
\begin{figure}
    \includegraphics[width=1.0\textwidth]{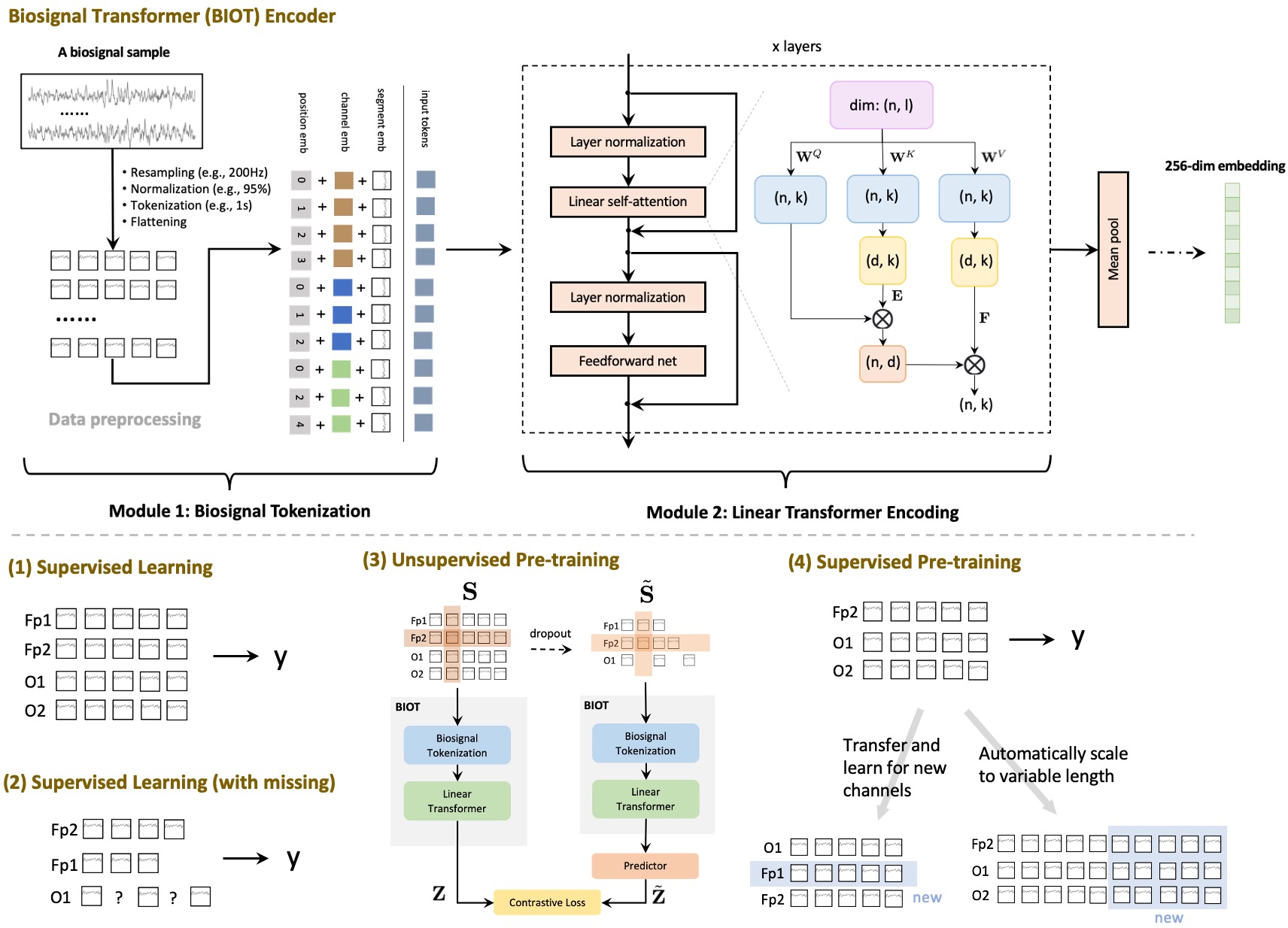}
    \caption{Biosignal Transformer (\method). (Upper) Given a new data sample, we initially perform data preprocessing (resampling, normalization, tokenization, and flattening) to create a biosignal "sentence" using the {\bf biosignal tokenization module}. We then learn the complex interactions within the "sentence" through the {\bf linear transformer module}. (Lower) \method encoder is versatile, enabling supervised learning on complete data, data with missing values, and pre-training and fine-tuning across diverse data formats and tasks.}
    \label{fig:bst}
\end{figure}

\vspace{-2mm}
As shown in Figure~\ref{fig:bst}, our \method encoder cascades two modules: (i) the {\bf biosignal tokenization} module that tokenizes an arbitrary biosignal (variable lengths, different channels, and missing values) into a "sentence" structure. This design can potentially enable previous language modeling techniques \citep{devlin2018bert,liu2019roberta,openai2023gpt4} to empower the current biosignal models; (ii) a {\bf linear transformer} module  that captures complex token interactions within the "sentence" while maintaining linear complexity. After that, we also discuss the application of \method in different real-world settings.

\subsection{Module 1: Biosignal Tokenization} \label{sec:sentencization}
\vspace{-1mm}

\noindent {\bf Motivation.} The goal of this paper is to model heterogeneous biosignals (e.g., EEG samples with different channels for different tasks) with a unified encoding model. For example, common EEG samples \citep{lopez2015automated}, such as those for seizure detection, are recorded at 256Hz in the international 10-20 system \footnote{https://en.wikipedia.org/wiki/10-20\_system\_(EEG)} for 10-second long~\citep{Klem1999TheTE}. With standard 16 montage editing, the sample is essentially a multi-channel time-series, represented as a matrix of size (16, 2560). However, format mismatch may prevent the applications on other similar data, such as {\bf different sampling rate} (e.g., 200Hz vs 256Hz) \citep{jing2023development}, {\bf mismatched channels} (i.e., different datasets have their own novel channels), {\bf variable recording duration} (i.e., 30s per sample vs. 10s) \citep{zhang2022self}, {\bf missing segments} (i.e., part of the recording is damaged due to device error). Thus, existing models may fail to utilize the mismatched data from different datasets. 

\begin{figure}
    \includegraphics[width=0.96\textwidth]{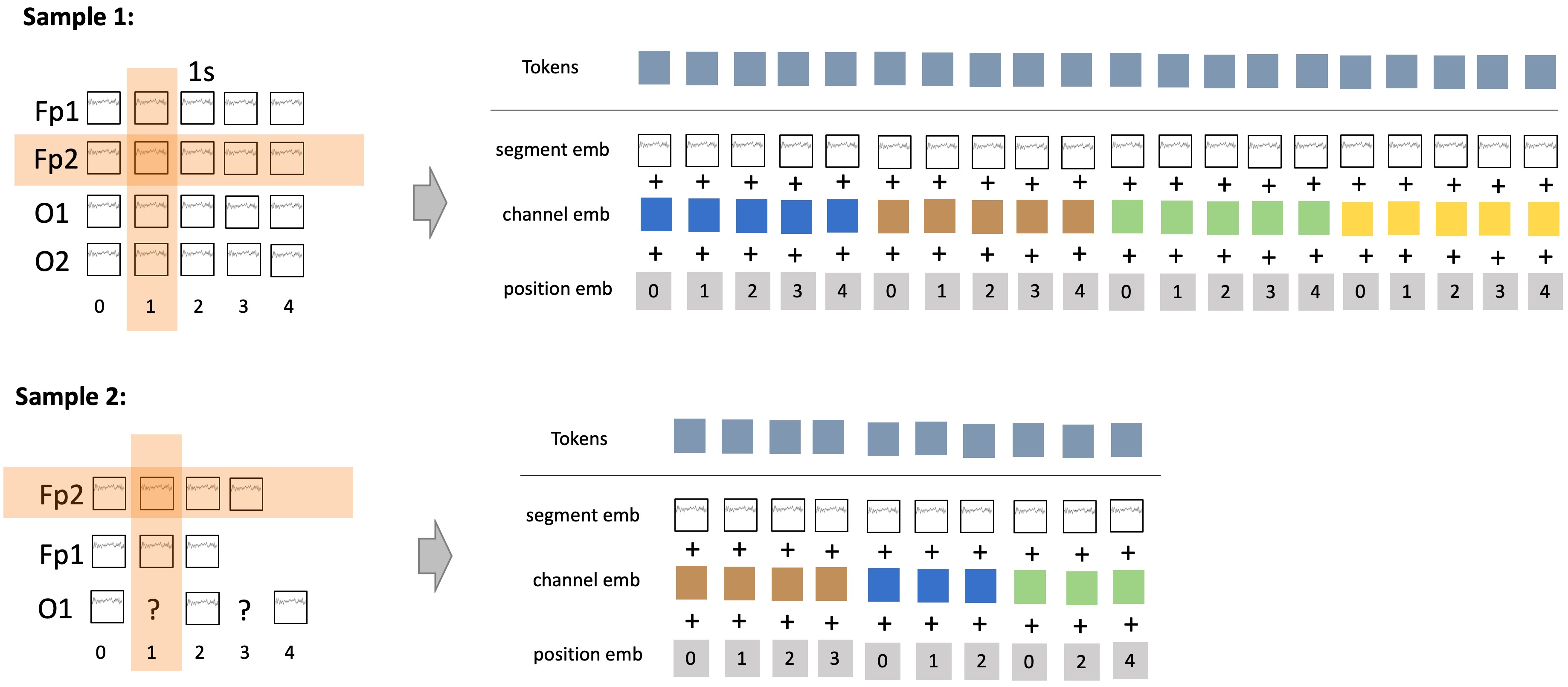}
    \caption{Biosignal Tokenization (no overlap in the examples). {\bf Sample 1} has four channels (Fp1, Fp2, O1, and O2) for 5 seconds. We tokenize each channel into segments and then parameterize these 20 segments with three embeddings. 
    On the right, we use different colors to represent the channels (\textcolor{blue}{blue}-Fp1, \textcolor{brown}{brown}-Fp2, \textcolor{green!50!gray}{green}-O1, and \textcolor{yellow!70!orange}{yellow}-O2). {\bf Sample 2} has mismatched channels (no O2), variable lengths (Fp1 and Fp2 are shorter and flipped), and missing values (in O1).
    Using our method, we can still tokenize Sample 2 in a comparable "sentence".}
    \label{fig:sentence}
    \vspace{-2mm}
\end{figure}

Our \method solves the above challenges by the following steps. Illustrations are shown in Figure~\ref{fig:sentence}. Assume the multi-channel biosignal as $\bfS\in\RR^{I\times J}$ (use a complete sample for the ease of notation). 
\vspace{-2mm}
\begin{itemize}
    \item {\bf Resampling.} We first resample all data to the same rate (denoted by $r\in\RR^+$, such as 200Hz) by linear interpolation. The consistent rate could be selected following clinical knowledge of a certain biosignal. For example, the highest frequency of interest in both EEG and ECG signals is commonly around 100 Hz, and thus 200 Hz or 250 Hz can be suitable for typical EEG or ECG applications, according to Nyquist-Shannon sampling theorem \citep{nyquist1928certain,shannon1949communication}. 
    \item {\bf Normalization}: To alleviate the unit difference and amplitude mismatch across different channels and datasets, we use the 95-percentile of the absolute amplitude to normalize each channel. Formally, each channel $\bfS[i]$ is normalized by $\frac{\bfS[i]}{\percentile(|\bfS[i]|,~95\%)}$.
    \item {\bf Tokenization}: For handling length variation, we tokenize the recording of each channel into $t$-second tokens, and neighboring tokens can have overlaps of $p$ seconds ($p,t\in\RR^+$ and $p<t$, e.g., $t=1$ and $p=0.5$). Thus, the $k$-th token ($k=1,2,3,...$) in the $i$-th channel can be represented by the slicing notation $\bfS[i,(t-p)(k-1):(t-p)(k-1)+t]$. The number of tokens per channel is limited by the inequality: $(t-p)(k-1)+t\leq J$. Here, the overlap $p$ is essential to maintain the temporal information for shorter signals. For example, if the length of signal is $J=3$, a configuration of $t=1,p=0$ will only generate 3 tokens for each channel, while a configuration of $t=1,p=0.5$ gives 5 tokens per channel.
    In cases of missing values, we drop the corresponding tokens directly (as shown in {\bf Sample 2} Figure~\ref{fig:sentence}). Note that, our tokenization applies to each channel, separately, which is different from previous works \citep{biswal2018expert,almutairi2021detection,du2022eeg} that split all channels together (which cannot work on {\bf Sample 2}).  
    \item {\bf Flattening}: We finally flatten tokens from all channels into a consistent "sentence".
\end{itemize}
\vspace{-1mm}
The above steps are non-parametric. To study the effect of sampling rate $r$, token length $t$, and overlap $p$, we provide ablation studies and insights in Appendix~\ref{sec:ablation-study}. In the following, we design the token embedding for the biosignal "sentence", which combines information from three aspects.
\vspace{-1mm}
\begin{itemize}
    \item {\bf Segment embedding.} We learn the segment embedding from a spectral perspective by first extracting an energy vector for each token $\bfS[i,(t-p)(k-1):(t-p)(k-1)+t]$ based on all frequency bands. This step is enabled by fast fourier transform (FFT). A fully connected network (FCN) is then applied on the energy vector to obtain the segment embedding.
    \item {\bf Channel embedding (spatial).} We learn an embedding table for all different channels and add the corresponding channel embedding to the token. Each color represents one channel in Figure~\ref{fig:sentence}.
    \item {\bf Positional embedding (temporal).} In biosignals, the segment order within the channel captures temporal information. We thus add {\em relative} positional embedding to the final toke embedding by using the sinusoidal and cosine functions, which does not need learnable parameters. 
\end{itemize}

\vspace{-1mm}
We denote the final tokenzied biosignal "sentence" as $\bfX\in\RR^{N\times l}$ where $N$ is the number of tokens and $l$ is the dimension of token embedding. In Figure~\ref{fig:sentence}, the marked \textcolor{gray!20!orange}{orange area} indicates the spatial- or temporal-relevant tokens w.r.t. the current token (i.e., same time step or channel). Our token embeddings can effectively capture the segment features as well as the spatio-temporal features.

\vspace{-2mm}
\subsection{Module 2: Linear transformer} \label{sec:linformer}
\vspace{-1mm}
\noindent {\bf Transformer with linear complexity for long biosignal "sentence".} Next, we want to leverage the Transformer model \citep{vaswani2017attention} for learning the "sentence" embedding. However, biosignals usually have many channels, which may lead to long "sentences". For example, the "sentence" of a 64-channel EEG signal for 20 seconds (without overlaps $p=0$) can have $64\times20 = 1280$ tokens, and longer with the overlaps. Given that the original Transformer model is known to have quadratic complexity in both time and space, we adopt the linear attention mechanism \citep{wang2020linformer,katharopoulos-et-al-2020} for biosignal learning applications.

Formally, let us assume $\bfW^K, \bfW^V,\bfW^Q\in\RR^{l\times k}$ be the key, value, and query matrices. Our self-attention module uses a rank-$d$ approximation for the softmax attention ($N\times N$) by reduced-rank parameter matrices $\bfE^\top\in\RR^{N\times d},\bfF\in\RR^{d\times N}$ (where $d\ll N$). The output $\bfH\in\RR^{N\times k}$ is,
\begin{align}
    \bfH = ~&\attention(\bfX\bfW^Q, \textcolor{purple}{\bfE}\bfX\bfW^K, \textcolor{purple}{\bfF}\bfX\bfW^V) \\
        = ~&\underbrace{\softmax\left(\frac{(\bfX\bfW^Q)(\textcolor{purple}{\bfE}\bfX\bfW^K)^\top}{\sqrt{k}}\right)}_{N\times d}\cdot \underbrace{\textcolor{purple}{\bfF}\bfX\bfW^V}_{d\times l}.
\end{align}
The main components in our linear transformer module are one linear self-attention layer and one fully connected network. To enable stable training, we add layer normalization \citep{ba2016layer}, residual connection \citep{he2016deep}, and dropout \citep{srivastava2014dropout} right before each component (see Figure~\ref{fig:bst}), which greatly accelerates the convergence and improves the final performance.

\noindent {\bf \method Encoder.} An illustration of our proposed \method encoder is shown in Figure~\ref{fig:bst} (upper), which comprises the {\bf biosignal tokenization} module and multiple blocks of {\bf linear transformer} modules. We obtain the final biosignal "sentence" embedding by a mean pooling step over all tokens. Note that appending a classification [CLS] token at the beginning of the "sentence" (after Module 1) is also a common option. However, we find it yields a slightly worse performance in our application, and thus we use mean pooling in the experiments.

\vspace{-1mm}
\subsection{Biosignal Learning in the Wild}
\vspace{-1mm}
Our proposed \method encoder can be applied in various real-world biosignal applications illustrated in the lower part of Figure~\ref{fig:bst}). These applications include (1) standard supervised learning, (2) learning with missing channels or segments, (3) \& (4) pre-training on one or more datasets, and fine-tuning on other similar datasets with different input formats.

\noindent {\bf (1) Supervised Learning} is the most common setting in the previous literature \citep{jing2023development,biswal2018expert}. With the \method encoder, we finally apply an exponential linear unit (ELU) activation \citep{clevert2015fast} and a linear layer for classification tasks.
\vspace{-1mm}

\noindent {\bf (2) Supervised Learning (with missing).} Many real biosignal data have mismatched channels, missing segments, and variable lengths, which prevents the applications of existing models \citep{jing2023development,song2021transformer}. Flexible as our model is, \method can be applied in this setting with the same model structure as in {\bf (1)}.
\vspace{-1mm}

\noindent {\bf (3) Unsupervised Pre-training.} We can jointly pre-train a general-purpose \method encoder on multiple large unlabeled datasets. In the experiments, we pre-train an unsupervised encoder using 5 million resting EEG samples (16 channels, 10s, 200Hz) and 5 million sleep EEG samples (2 channels, 30s, 125Hz), which is later utilized to improve various downstream tasks. 
\vspace{-1mm}

For the unsupervised pre-training, we take the following steps (a diagram is shown in Figure~\ref{fig:bst}). 
\vspace{-1mm}
\begin{itemize}
    \item Assume $\bfS$ is the original biosignal. We first randomly dropout part of its channels and dropout part of the tokens from the remaining channels, resulting in a perturbed signal $\tilde{\bfS}$. 
    \vspace{-1mm}
    \item We then obtain the embeddings of $\bfS$ and $\tilde{\bfS}$ by the same \method encoder. To form the objective, we want to predict the embedding of the original signal by the perturbed signal. Thus, an additioanl predictor (i.e., two-layer neural network) is appended for the perturbed signal following \citep{grill2020bootstrap}. We use $\bfZ$ and $\tilde{\bfZ}$ to denote the real embedding of $\bfS$ and predicted embedding from $\tilde{\bfS}$.
    \begin{align}
    \bfZ =~ \method(\bfS),~~\tilde{\bfZ} =~ \predictor(\method(\tilde{\bfS})).
    \end{align}
    \vspace{-2mm}
    \item Finally, contrastive loss \citep{he2020momentum,chen2020simple} is used on $\bfS$ and $\tilde{\bfS}$ to form the objective.
    \begin{align}
    \calL =~\crossentropy\left(\softmax\left(\langle\bfZ,\tilde{\bfZ}^\top\rangle/T\right),\bfI\right).
    \vspace{-2mm}
\end{align}
\end{itemize}
Here, $T$ represents the temperature ($T=0.2$ throughout the paper) and $\bfI$ is an identity matrix. In the implementation, we also apply sample-wise L2-normalization on both $\bfZ$ and $\tilde{\bfZ}$ before softmax.

\noindent {\bf (4) Supervised Pre-training} aims to pre-train a model by supervised learning on one task and then generalize and fine-tune the encoder on a new task. The goal is to transfer knowledge among different datasets and gain improvements on the new task compared to training from scratch. Our \method model allows the new datasets to have mismatched channels and different lengths.

\vspace{-1mm}
\section{Experiments}
\vspace{-2mm}
This section shows the strong performance of \method on several EEG, ECG and sensory datasets. Section~\ref{sec:supervised-learning}, \ref{sec:learning-with-missing} compare \method with baselines on {\bf supervised learning} and {\bf learning with missing} settings. Section~\ref{sec:unsupervised-pre-train}, \ref{sec:supervised-pre-train}, \ref{sec:ultimate} show the that \method can be flexibly {\bf pre-trained on other datasets} (supervised or unsupervised) to improve the current task with different sample formats.
We released the codebase and the pre-trained models in GitHub \footnote{https://github.com/ycq091044/BIOT}.
\vspace{-1mm}
\subsection{Experimental Setups}
\vspace{-1mm}
\noindent {\bf Biosignal Datasets.} We consider the following datasets in the evaluation: (i) {\bf SHHS} \citep{zhang2018national,quan1997sleep} is a large sleep EEG corpus from patients aged 40 years and older. (ii) {\bf PREST} is a large unlabeled proprietary resting EEG dataset; (iii) {\bf Cardiology} \citep{alday2020classification} is a collection of five ECG datasets (initially contains six, but we exclude the PTB-XL introduced below). (iv) The {\bf CHB-MIT} database \citep{shoeb2009application} is collected from pediatric patients for epilepsy seizure detection. (v) {\bf IIIC Seizure} dataset is from \cite{ge2021deep,jing2023development} for detecting one of the six ictal-interictal-injury-continuum (IIIC) seizure patterns (OTH, ESZ, LPD, GPD, LRDA, GRDA); (vi) TUH Abnormal EEG Corpus ({\bf TUAB}) \citep{lopez2015automated} is an EEG dataset that has been annotated as normal or abnormal; (vii) TUH EEG Events ({\bf TUEV}) \citep{harati2015improved} is a corpus of EEG that contains annotations of EEG segments as one of six event types: spike and sharp wave (SPSW), generalized periodic epileptiform discharges (GPED), periodic lateralized epileptiform discharges (PLED), eye movement (EYEM), artifact (ARTF) and background (BCKG); (viii) {\bf PTB-XL} \citep{wagner2020ptb} is an ECG dataset with 12-lead recordings for diagnosis prediction, and we used it for arrhythmias phenotyping in this paper; (ix) {\bf {HAR}} \citep{anguita2013public} is a human action recognition dataset using smartphone accelerometer and gyroscope data.

\begin{table}[h!]
\centering
\caption{Dataset Statistics}
\resizebox{1.02\textwidth}{!}{\begin{tabular}{c|ccccccc} 
\toprule 
  {\bf Datasets} & {\bf Type (subtype)} & {\bf \# Recordings} & {\bf Rate} & {\bf Channels} & {\bf Duration} & {\bf \# Sample} & {\bf Tasks}  \\
  \midrule
  SHHS & EEG (sleep) & 5,445 & 125Hz & C3-A2, C4-A1 & 30 seconds & 5,093,522 & Unsupervised pre-training \\
  PREST & EEG (resting) & 6,478 & 200Hz & 16 montages & 10 seconds & 5,110,992 & Unsupervised pre-training \\
  Cardiology & ECG & 21,264 & 500Hz & 6 or 12 ECG leads & 10 seconds & 495,970 & Unsupervised pre-training \\
  \midrule
  CHB-MIT & EEG (resting) & 686 & 256Hz & 16 montages & 10 seconds & 326,993 & Binary (seizure or not) \\
  IIIC Seizure & EEG (resting) & 2,702 & 200Hz & 16 montages & 10 seconds & 165,309 & Multi-class (6 seizure types) \\
  TUAB & EEG (unknown) & 2,339 & 256Hz & 16 montages & 10 seconds & 409,455 & Binary (abnormal or not) \\ 
  TUEV & EEG (both) & 11,914 & 256Hz & 16 montages & 5 seconds & 112,491 & Multi-class (6 event types) \\
  \midrule
  PTB-XL & ECG & 21,911 & 500Hz & 12 ECG leads & 5 seconds & 65,511 & Binary (arrhythmias or not) \\
  HAR & Wearable sensors & 10,299 & 50Hz & 9 coordinates & 2.56 seconds & 10,299 & Multi-class (6 actions) \\
\bottomrule
\end{tabular}}
\label{tb:datasets} 
\vspace{-1mm}
\end{table}

\noindent {\bf Dataset Processing.} The first three datasets are used entirely for unsupervised pre-training. The next four datasets are used for supervised learning, and we used the common 16 bipolar montage channels in the international 10-20 system. For CHB-MIT (containing 23 patients), we first use patient 1 to 19 for training, 20,21 for validation, and 22,23 for test. Then, we flip the validation and test sets and conduct the experiments again. We report the average performance on these two settings. For IIIC seizure, we divide patient groups into training/validation/test sets by 60\%:20\%:20\%. For TUAB and TUEV, the training and test separation is provided by the dataset. We further divide the training patients into training and validation groups by 80\%:20\%. For PTB-XL, we divide patient groups into training/validation/test sets by 80\%:10\%:10\%. The train and test set of HAR is provided, and we further divide the test patients into validation/text by 50\%:50\%. For all the datasets, after assigning the patients to either training, validation, or test groups, we will further split the patient's recording to samples, and the sample duration accords to the annotation files.
The dataset statistics can be found in Table~\ref{tb:datasets}, and we provides more descriptions and processing details in Appendix~\ref{sec:datasets-details}.

\noindent {\bf Baseline.} We consider the following representative models: (i) {\bf SPaRCNet} \citep{jing2023development} is a 1D-CNN based model with dense residual connections, more advanced than the popular ConvNet \citep{schirrmeister2017deep}, CSCM \citep{sakhavi2018c2cm}; (ii) {\bf ContraWR}'s \citep{yang2021self} encoder model first transforms the biosignals into multi-channel spectrogram and then uses 2D-CNN based ResNet \citep{he2016deep}; (iii) {\bf CNN-Transformer} \citep{peh2022transformer} is superior to CNN-LSTM models \citep{zhang2019cnnlstm}; (iv) {\bf FFCL} \citep{li2022motor} combines embeddings from CNN and LSTM encoders for feature fusion; (v) {\bf ST-Transformer} \cite{song2021transformer} proposes an multi-level EEG transformer for learning spatial (S) and temporal (T) features simultaneously, empirically better than EEGNet \cite{lawhern2018eegnet}. Our \method model trained from scratch is denoted by (vanilla).

\noindent {\bf Environments and Settings.} The experiments are implemented by Python 3.9.12, Torch 1.13.1+cu117, Pytorch-lightning 1.6.4 on a Linux server with 512 GB memory, 128-core CPUs and eight RTX A6000 GPUs. All the models are optimized on training set and evaluated on the test set. The best model and hyperparameter combinations are selected based on the validation set. For Table~\ref{tb:eeg-task1} and Table~\ref{tb:ecg-har}, we obtain five sets of results with different random seeds and report the mean and standard deviation values. For Figure~\ref{fig:missing} and Figure~\ref{fig:transfer}, we report the results under three random seeds. More experimental and implementation details can refer to Appendix~\ref{sec:experiment-details}.

\subsection{Setting (1) - standard supervised learning} \label{sec:supervised-learning}
This section shows that \method is comparable or better than baselines in the supervised learning settings.
\begin{itemize}
    \vspace{-1mm}
    \item {\bf Four EEG Tasks.} Both CHB-MIT and TUAB are designed to predict binary output, and we use binary cross entropy (BCE) for TUAB and the focal loss \citep{lin2017focal} for CHB-MIT due to its imbalances (around 0.6\% positive ratio in training set). We use balanced accuracy (Balanced Acc.), area under precision-recall curve (AUC-PR) and AUROC as the metrics. Both IIIC Seizure and TUEV are multi-class classification tasks with cross entropy loss. We employ Balanced Acc., Cohen's Kappa, and Weighted F1 as the multi-class evaluation. To save space, we only show the performance on CHB-MIT and IIIC Seizure in Table~\ref{tb:eeg-task1} and move the other two to Appendix~\ref{sec:eeg-task2}. 
    \item {\bf ECG and Sensory Tasks.} PTB-XL is formulated as a binary classification on detecting arrhythmias phenotypes. We use the BCE loss and binary evaluation metrics. HAR (classifying actions) uses the cross entropy loss and is evaluated by multi-class metrics. Results are reported in Table~\ref{tb:ecg-har}.
     \vspace{-1mm}
\end{itemize} 
Table~\ref{tb:eeg-task1} and \ref{tb:ecg-har} show that our model has superior performance over baselines in most tasks, especially on CHB-MIT, IIIC Seizure, and HAR. The reason might be that the frequency features are more useful in these three datasets as our \method extracts the main features from spectral perspective. SPaRCNet is a strong model among all the baselines except on the CHB-MIT task. The model might be vulnerable in the imbalanced classification setting even with the focal loss. 
The pre-training models at the end of the tables will be introduced and explained in Section~\ref{sec:unsupervised-pre-train},~\ref{sec:ultimate}.
\begin{table}[h!]
\centering
\caption{EEG classification tasks (Results of TUAB and TUEV are in Appendix~\ref{sec:eeg-task2})}
\resizebox{1.02\textwidth}{!}{\begin{tabular}{l|ccc|ccc} 
\toprule 
  \multirow{2}{*}{\bf Models} & \multicolumn{3}{c|}{\bf CHB-MIT (seizure detection)} & \multicolumn{3}{c}{\bf IIIC Seizure (seizure type classification)} \\
  \cmidrule{2-7}
  & Balanced Acc. & AUC-PR & AUROC & Balanced Acc. & Cohen’s Kappa & Weighted F1 \\
  \midrule
SPaRCNet \citep{jing2023development}       & 0.5876 $\pm$ 0.0191 & 0.1247 $\pm$ 0.0119 & 0.8143 $\pm$ 0.0148 & 0.5546 $\pm$ 0.0161 & 0.4679 $\pm$ 0.0228 & 0.5569 $\pm$ 0.0184\\
ContraWR \citep{yang2021self}       & 0.6344 $\pm$ 0.0002 & 0.2264 $\pm$ 0.0174 & 0.8097 $\pm$ 0.0114 & 0.5519 $\pm$ 0.0058 & 0.4623 $\pm$ 0.0148 & 0.5486 $\pm$ 0.0137\\
CNN-Transformer \citep{peh2022transformer} & 0.6389 $\pm$ 0.0067 & 0.2479 $\pm$ 0.0227 & {\bf 0.8662} $\pm$ 0.0082 & 0.5476 $\pm$ 0.0103 & 0.4481 $\pm$ 0.0139 & 0.5346 $\pm$ 0.0127 \\
FFCL  \citep{li2022motor}     & 0.6262 $\pm$ 0.0104 & 0.2049 $\pm$ 0.0346 & 0.8271 $\pm$ 0.0051 & 0.5617 $\pm$ 0.0117 & 0.4704 $\pm$ 0.0130 & 0.5617 $\pm$ 0.0171 \\
ST-Transformer \citep{song2021transformer} & 0.5915 $\pm$ 0.0195 & 0.1422 $\pm$ 0.0094 & 0.8237 $\pm$ 0.0491 & 0.5423 $\pm$ 0.0056 & 0.4492 $\pm$ 0.0056 & 0.5440 $\pm$ 0.0014 \\
(Vanilla) \method     & {\bf 0.6640 $\pm$ 0.0037} & {\bf 0.2573 $\pm$ 0.0088} & 0.8646 $\pm$ 0.0030 & {\bf 0.5762 $\pm$ 0.0034} & {\bf 0.4932 $\pm$ 0.0046} & {\bf 0.5773 $\pm$ 0.0031} \\
\midrule
Pretrained \method (PREST) & 0.6942 $\pm$ 0.0431 & 0.3072 $\pm$ 0.1187 & 0.8679 $\pm$ 0.0106 & 0.5787 $\pm$ 0.0066 & 0.4980 $\pm$ 0.0054 & 0.5828 $\pm$ 0.0049  \\
Pretrained \method (PREST+SHHS) & 0.6788 $\pm$ 0.0036 & {0.3090 $\pm$ 0.0003} & {0.8752 $\pm$ 0.0022} & \fbox{0.5800 $\pm$ 0.0004} & \fbox{0.5040 $\pm$ 0.0041} & \fbox{0.5878 $\pm$ 0.0015} \\
Pretrained \method (6 EEG datasets) & \fbox{0.7068 $\pm$ 0.0457} & \fbox{0.3277 $\pm$ 0.0460} & \fbox{0.8761 $\pm$ 0.0284} & 0.5779 $\pm$ 0.0087 & 0.4949 $\pm$ 0.0103 & 0.5737 $\pm$ 0.0088  \\
\bottomrule
\multicolumn{7}{l}{1. All models use the same training set of the task, while the pre-trained \method models are initially pre-trained on other data sources (see Section~\ref{sec:unsupervised-pre-train},~\ref{sec:ultimate}).} \\
\multicolumn{7}{l}{2. {\bf Bold} for the best model (trained from scratch) and \fbox{box} for the best pre-trained models. 
}
\end{tabular}}
\label{tb:eeg-task1} 
\vspace{-2mm}
\end{table}

\begin{table}[h!]
\centering
\caption{ECG and human activity sensory classification tasks}
\resizebox{1.0\textwidth}{!}{\begin{tabular}{l|ccc|ccc} 
\toprule 
  \multirow{2}{*}{\bf Models} & \multicolumn{3}{c|}{\bf PTB-XL (arrhythmias phenotype prediction)} & \multicolumn{3}{c}{\bf HAR (huamn action recognition)} \\
  \cmidrule{2-7}
  & Balanced Acc. & AUC-PR & AUROC & Balanced Acc. & Cohen’s Kappa & Weighted F1  \\
  \midrule
SPaRCNet \citep{jing2023development}       & 0.8275 $\pm$ 0.0047 & {\bf 0.9040 $\pm$ 0.0067} & {\bf 0.7550 $\pm$ 0.0073} & 0.9371 $\pm$ 0.0160 & 0.9236 $\pm$ 0.0189 & 0.9365 $\pm$ 0.0155 \\
ContraWR \citep{yang2021self}        & 0.6341 $\pm$ 0.0883 & 0.6795 $\pm$ 0.1083 & 0.4433 $\pm$ 0.1557 & 0.9068 $\pm$ 0.0164 & 0.8879 $\pm$ 0.0201 & 0.9055 $\pm$ 0.0182 \\
CNN-Transformer \citep{peh2022transformer} & 0.6650 $\pm$ 0.0459 & 0.7175 $\pm$ 0.0558 & 0.4996 $\pm$ 0.0936 & 0.8690 $\pm$ 0.0839 & 0.8273 $\pm$ 0.0953 & 0.8352 $\pm$ 0.1166 \\
FFCL \citep{li2022motor}          & 0.7034 $\pm$ 0.0052 & 0.7088 $\pm$ 0.0053 & 0.5127 $\pm$ 0.0051 & 0.8519 $\pm$ 0.0148 & 0.8216 $\pm$ 0.0177 & 0.8508 $\pm$ 0.0138  \\
ST-Transformer \citep{song2021transformer} & 0.7238 $\pm$ 0.0083 & 0.7775 $\pm$ 0.0153 & 0.6003 $\pm$ 0.0179 & 0.9336 $\pm$ 0.0063 & 0.9213 $\pm$ 0.0076 & 0.9337 $\pm$ 0.0068 \\
(Vanilla) \method &  {\bf 0.8315 $\pm$ 0.0008} & 0.8978 $\pm$ 0.0020 & 0.7493 $\pm$ 0.0167 & {\bf 0.9461 $\pm$ 0.0134} & {\bf 0.9351 $\pm$ 0.0160} & {\bf 0.9458 $\pm$ 0.0136}  \\
\midrule
Pretrained \method (Cardiology-6) & 0.8350 $\pm$ 0.0073 & 0.9128 $\pm$ 0.0094 & \fbox{0.7671 $\pm$ 0.0116} & /& /& / \\
Pretrained \method (Cardiology-12) & \fbox{0.8421 $\pm$ 0.0030} & \fbox{0.9221 $\pm$ 0.0075} & 0.7659 $\pm$ 0.0076 & /& /& / \\
\bottomrule
\multicolumn{7}{l}{* {\bf Bold} for the best model. All models use the same training set of the task. The Pretrained \method (Cardiology-6) and Pretrained \method (Cardiology-12)}\\
\multicolumn{7}{l}{~~~are pre-trained on Cardiology data (see Section~\ref{sec:unsupervised-pre-train}), and they do not apply to HAR data (due to different biosignal types).
}
\end{tabular}}
\label{tb:ecg-har} 
\end{table}

\subsection{Setting (2) - learning with missing channels and segments} \label{sec:learning-with-missing}
The section simulates the TUEV dataset to mimic the setting of {\em supervised learning with missing channels and segments} and show the strong performance of \method. We consider three missing cases:
\vspace{-1mm}
\begin{itemize}
    \item {\bf Missing segments}: Randomly mask out $a$ segments (each segment spans for 0.5 seconds), $a=0,1,2,3,4,5$ with equal probability. The segment masking is applied separately for each channel.
    \item {\bf Missing channels}: Randomly mask out $b$ channels, $b=0,1,2,3,4$ with equal probability. We assume that the masking will not alter the underlying labels (the same assumption for other cases).
    \item {\bf Missing both channels and segments}: Combining Case 2 \& 3 simultaneously.
\end{itemize}

\begin{figure} [t!]
    \includegraphics[width=0.97\textwidth]{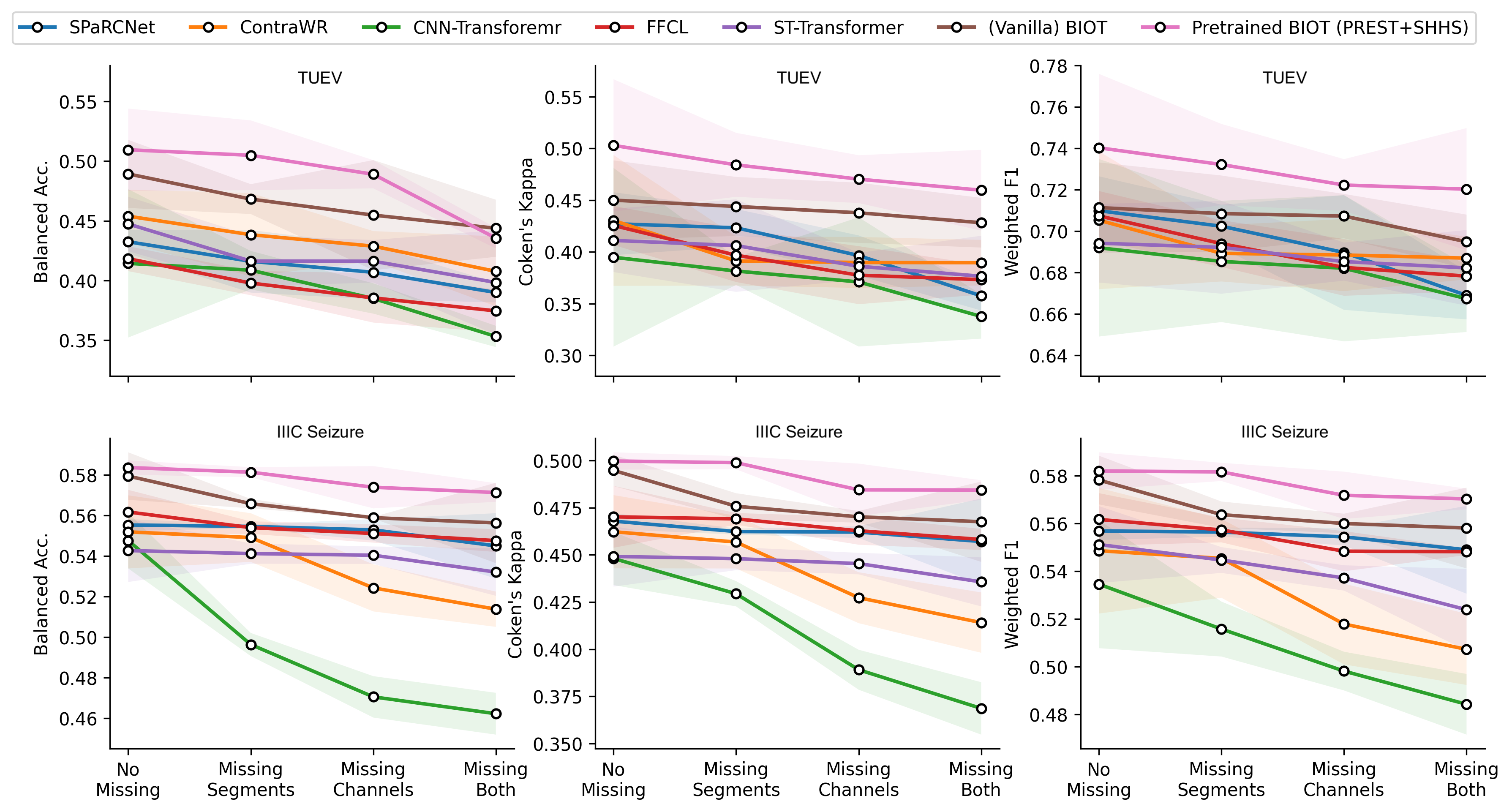}
    \caption{Supervised learning with missing channels or segments (on TUEV and IIIC Seizure)}
    \label{fig:missing}
\end{figure}

To enable the baseline models compatible with the setting, we use all zeros to impute the masked regions. The comparison is plotted in Figure~\ref{fig:missing}, which shows that (i) all models decrease the performance with more missings while \method and the pre-trained \method are less impacted (especially on Kappa and Weighted F1); (ii) "Missing channels" affects the performance more than "Missing segments", which makes sense as segment masking still preserves information from all channels.

\subsection{Setting (3) - unsupervised pre-training} \label{sec:unsupervised-pre-train}
In this section, we show that \method enables unsupervised pre-training on existing with various formats.
\begin{itemize}
    \item {\bf Pre-trained  (PREST)}: This model is pre-trained on 5 million resting EEG samples (PREST) with 2,048 as the batch size. We save the pre-trained model at the 100-th epoch.
    \item {\bf Pre-trained (PRESET+SHHS)}: This model is jointly pre-trained on 5M PREST and 5M SHHS EEG samples. Though two datasets have different sample formats, our model is able to encode them regardless. Also, we use 2048 as the batch size and save model at the 100-th epoch.
    \item {\bf Pre-trained (Cardiology-12)} is jointly pre-trained on raw data of five datasets in Cardiology corpus (details in Appendix~\ref{sec:datasets-details}). We use 1024 as batch size and save model at the 100-th epoch.
    \item {\bf Pre-trained (Cardiology-6)} is pre-trained similarly as Pre-trained (Cardiology-12), while we only utilize the first 6 ECG leads. By contast, Pre-trained (Cardiology-12) uses full 12 leads. 
\end{itemize}
We fine-tune the first two pre-trained EEG models on four EEG tasks and append the results to Table~\ref{tb:eeg-task1} (also in Appendix~\ref{sec:eeg-task2}). We fine-tune the last two pre-trained ECG models on PTB-XL datasets in Table~\ref{tb:ecg-har}. The results show that the pre-trained models greatly improves the final performance on the downstream tasks in Table~\ref{tb:eeg-task1} and Table~\ref{tb:ecg-har}.

\subsection{Setting (4) - supervised pre-training on other tasks} \label{sec:supervised-pre-train}
\vspace{-1mm}
This section shows that \method allows knowledge transfer from one task to another similar task with different sample formats. We pre-train on  the training set of CHB-MIT, IIIC Seizure, TUAB and fine-tunes on TUEV (which has 16 channels and 5s duration). All datasets use 200Hz sampling rate. We design three sets of configurations for the pre-trained datasets: {\bf Format (i)} uses the first 8 channels and 10s duration; {\bf Format (ii)} uses the full 16 channels but only the first 5s recording; {\bf Format (iii)} uses full 16 channels and full 10s recording. During fine-tuning, we then remove the prediction layers from these pre-trained model and add a new prediction layer to fit the TUEV dataset.

\begin{figure}[!th]
    \includegraphics[width=0.95\textwidth]{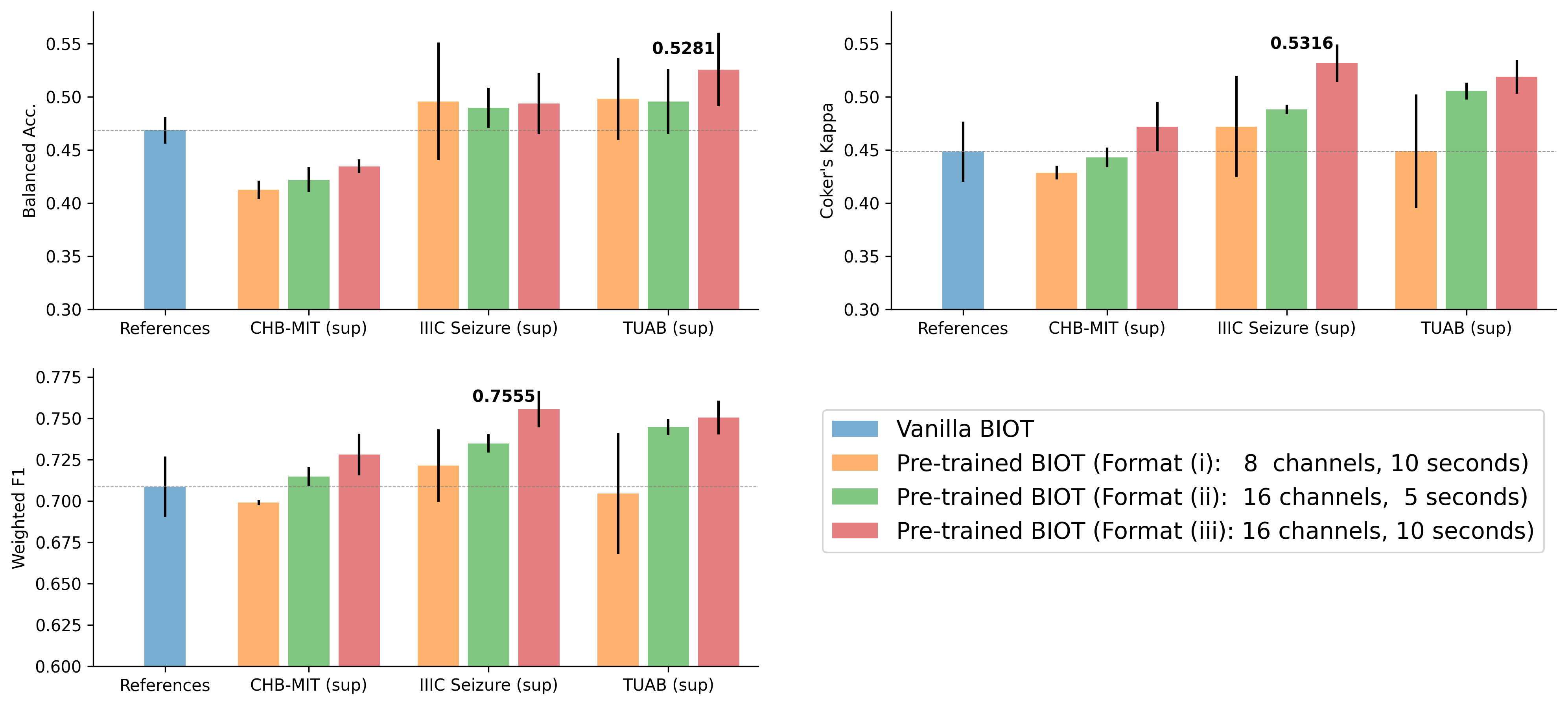}
    \vspace{-1mm}
    \caption{Fine-tuned on TUEV from different supervised pre-trained models ({best number in {\bf bold}}). Similar supervised fine-tuning analysis on CHB-MIT dataset is shown in Appendix~\ref{sec:chb-mit}.}
    \label{fig:transfer}
\end{figure}

The results are shown in Figure~\ref{fig:transfer} where we also add the vanilla \method (trained from scratch) for references. We find that (i) the model pre-trained on IIIC Seizure and TUAB are generally beneficial for the event classification task on TUEV. The reason might be that TUAB and TUEV are both recorded from Temple University and share some common information, while IIIC seizure and TUEV are both related to seizure detection and may share some latent patterns. (ii) More pre-training data will be beneficial in the downstream task. Though the pre-training configuration (16 channels, 5 seconds) aligns better with the TUEV data formats, the results show that configuration of (16 channels, 10 seconds) encodes longer duration and works consistently better. (iii) Compared to the TUEV results in Appendix~\ref{sec:eeg-task2}, we also find that oftentimes the supervised pre-training (e.g., on IIIC seizure or TUAB) can be more effective than unsupervised pre-training (e.g., on SHHS and PREST).


\vspace{-1mm}
\subsection{Pre-trained on all EEG datasets} \label{sec:ultimate}
\vspace{-1mm}
In this section, we show that \method can leverage all six EEG resources considered in the paper. We obtain a {\bf Pre-trained (6 EEG datasets)} model by loading the {Pre-trained (PREST+SHHS)} model and further train it on the training sets of CHB-MIT, IIIC Seizure, TUAB, and TUEV. We add separate classification layers for four tasks. Essentially, this model is pre-trained on all six EEG datasets. To use the model, we still fine-tune it on the training set of downstream tasks and append the results to Table~\ref{tb:eeg-task1} and Appendix~\ref{sec:eeg-task2}. Apparently, {Pre-trained (six EEG datasets)} outperforms the vanilla \method and is generally better than the unsupervised and the supervised pre-trained \method.

\vspace{-2mm}
\section{Conclusion}
\vspace{-3mm}
This paper proposes a new biosignal transformer model (\method) that learns embeddings for biosignals with variable lengths, channels and missing values. \method can enable effective knowledge transfer across different data and allow joint training on multiple sources. We conduct extensive evaluations on two large EEG corpus (5M each) for unsupervised pre-training, and several EEG, ECG, human action sensory datasets for supervised learning. The results show that our BIOT outperforms strong baselines in standard supervised learning and can effectively handle the learning settings with missing values. The pre-trained \method models also show significant improvements on various downstream classification tasks. In the end, we hope our work can inspire more follow-up researches of large foundational models for biosignals.

\bibliographystyle{apalike}
\bibliography{reference}

\begin{thebibliography}{}

\bibitem[Alday et~al., 2020]{alday2020classification}
Alday, E. A.~P., Gu, A., Shah, A.~J., Robichaux, C., Wong, A.-K.~I., Liu, C.,
  Liu, F., Rad, A.~B., Elola, A., Seyedi, S., et~al. (2020).
\newblock Classification of 12-lead ecgs: the physionet/computing in cardiology
  challenge 2020.
\newblock {\em Physiological measurement}, 41(12):124003.

\bibitem[Almutairi et~al., 2021]{almutairi2021detection}
Almutairi, H., Hassan, G.~M., and Datta, A. (2021).
\newblock Detection of obstructive sleep apnoea by ecg signals using deep
  learning architectures.
\newblock In {\em 2020 28th European signal processing conference (EUSIPCO)},
  pages 1382--1386. IEEE.

\bibitem[Anguita et~al., 2013]{anguita2013public}
Anguita, D., Ghio, A., Oneto, L., Parra, X., Reyes-Ortiz, J.~L., et~al. (2013).
\newblock A public domain dataset for human activity recognition using
  smartphones.
\newblock In {\em Esann}, volume~3, page~3.

\bibitem[Ba et~al., 2016]{ba2016layer}
Ba, J.~L., Kiros, J.~R., and Hinton, G.~E. (2016).
\newblock Layer normalization.
\newblock {\em arXiv preprint arXiv:1607.06450}.

\bibitem[Bahador et~al., 2021]{bahador2021reconstruction}
Bahador, N., Jokelainen, J., Mustola, S., and Kortelainen, J. (2021).
\newblock Reconstruction of missing channel in electroencephalogram using
  spatiotemporal correlation-based averaging.
\newblock {\em Journal of Neural Engineering}, 18(5):056045.

\bibitem[Biswal et~al., 2018]{biswal2018expert}
Biswal, S., Sun, H., Goparaju, B., Westover, M.~B., Sun, J., and Bianchi, M.~T.
  (2018).
\newblock Expert-level sleep scoring with deep neural networks.
\newblock {\em Journal of the American Medical Informatics Association},
  25(12):1643--1650.

\bibitem[Chen et~al., 2020]{chen2020simple}
Chen, T., Kornblith, S., Norouzi, M., and Hinton, G. (2020).
\newblock A simple framework for contrastive learning of visual
  representations.
\newblock In {\em International conference on machine learning}, pages
  1597--1607. PMLR.

\bibitem[Clevert et~al., 2015]{clevert2015fast}
Clevert, D.-A., Unterthiner, T., and Hochreiter, S. (2015).
\newblock Fast and accurate deep network learning by exponential linear units
  (elus).
\newblock {\em arXiv preprint arXiv:1511.07289}.

\bibitem[Cui et~al., 2020]{cui2020eeg}
Cui, H., Liu, A., Zhang, X., Chen, X., Wang, K., and Chen, X. (2020).
\newblock Eeg-based emotion recognition using an end-to-end regional-asymmetric
  convolutional neural network.
\newblock {\em Knowledge-Based Systems}, 205:106243.

\bibitem[Dar et~al., 2020]{dar2020cnn}
Dar, M.~N., Akram, M.~U., Khawaja, S.~G., and Pujari, A.~N. (2020).
\newblock Cnn and lstm-based emotion charting using physiological signals.
\newblock {\em Sensors}, 20(16):4551.

\bibitem[Devlin et~al., 2018]{devlin2018bert}
Devlin, J., Chang, M.-W., Lee, K., and Toutanova, K. (2018).
\newblock Bert: Pre-training of deep bidirectional transformers for language
  understanding.
\newblock {\em arXiv preprint arXiv:1810.04805}.

\bibitem[Dosovitskiy et~al., 2020]{dosovitskiy2020image}
Dosovitskiy, A., Beyer, L., Kolesnikov, A., Weissenborn, D., Zhai, X.,
  Unterthiner, T., Dehghani, M., Minderer, M., Heigold, G., Gelly, S., et~al.
  (2020).
\newblock An image is worth 16x16 words: Transformers for image recognition at
  scale.
\newblock {\em arXiv preprint arXiv:2010.11929}.

\bibitem[Du et~al., 2022]{du2022eeg}
Du, Y., Xu, Y., Wang, X., Liu, L., and Ma, P. (2022).
\newblock Eeg temporal--spatial transformer for person identification.
\newblock {\em Scientific Reports}, 12(1):14378.

\bibitem[Ge et~al., 2021]{ge2021deep}
Ge, W., Jing, J., An, S., Herlopian, A., Ng, M., Struck, A.~F., Appavu, B.,
  Johnson, E.~L., Osman, G., Haider, H.~A., et~al. (2021).
\newblock Deep active learning for interictal ictal injury continuum eeg
  patterns.
\newblock {\em Journal of neuroscience methods}, 351:108966.

\bibitem[Gong et~al., 2021]{gong2021ast}
Gong, Y., Chung, Y.-A., and Glass, J. (2021).
\newblock Ast: Audio spectrogram transformer.
\newblock {\em arXiv preprint arXiv:2104.01778}.

\bibitem[Greco et~al., 2021]{greco2021acute}
Greco, A., Valenza, G., L{\'a}zaro, J., Garz{\'o}n-Rey, J.~M., Aguil{\'o}, J.,
  De-la Camara, C., Bail{\'o}n, R., and Scilingo, E.~P. (2021).
\newblock Acute stress state classification based on electrodermal activity
  modeling.
\newblock {\em IEEE Transactions on Affective Computing}.

\bibitem[Grill et~al., 2020]{grill2020bootstrap}
Grill, J.-B., Strub, F., Altch{\'e}, F., Tallec, C., Richemond, P.,
  Buchatskaya, E., Doersch, C., Avila~Pires, B., Guo, Z., Gheshlaghi~Azar, M.,
  et~al. (2020).
\newblock Bootstrap your own latent-a new approach to self-supervised learning.
\newblock {\em Advances in neural information processing systems},
  33:21271--21284.

\bibitem[Harati et~al., 2015]{harati2015improved}
Harati, A., Golmohammadi, M., Lopez, S., Obeid, I., and Picone, J. (2015).
\newblock Improved eeg event classification using differential energy.
\newblock In {\em 2015 IEEE Signal Processing in Medicine and Biology Symposium
  (SPMB)}, pages 1--4. IEEE.

\bibitem[He et~al., 2020]{he2020momentum}
He, K., Fan, H., Wu, Y., Xie, S., and Girshick, R. (2020).
\newblock Momentum contrast for unsupervised visual representation learning.
\newblock In {\em Proceedings of the IEEE/CVF conference on computer vision and
  pattern recognition}, pages 9729--9738.

\bibitem[He et~al., 2016]{he2016deep}
He, K., Zhang, X., Ren, S., and Sun, J. (2016).
\newblock Deep residual learning for image recognition.
\newblock In {\em Proceedings of the IEEE conference on computer vision and
  pattern recognition}, pages 770--778.

\bibitem[Isin and Ozdalili, 2017]{isin2017cardiac}
Isin, A. and Ozdalili, S. (2017).
\newblock Cardiac arrhythmia detection using deep learning.
\newblock {\em Procedia computer science}, 120:268--275.

\bibitem[Jing et~al., 2018]{jing2018rapid}
Jing, J., d’Angremont, E., Zafar, S., Rosenthal, E.~S., Tabaeizadeh, M.,
  Ebrahim, S., Dauwels, J., and Westover, M.~B. (2018).
\newblock Rapid annotation of seizures and interictal-ictal continuum eeg
  patterns.
\newblock In {\em 2018 40th Annual International Conference of the IEEE
  Engineering in Medicine and Biology Society (EMBC)}, pages 3394--3397. IEEE.

\bibitem[Jing et~al., 2023]{jing2023development}
Jing, J., Ge, W., Hong, S., Fernandes, M.~B., Lin, Z., Yang, C., An, S.,
  Struck, A.~F., Herlopian, A., Karakis, I., et~al. (2023).
\newblock Development of expert-level classification of seizures and rhythmic
  and periodic patterns during eeg interpretation.
\newblock {\em Neurology}.

\bibitem[Jing et~al., 2020]{jing2020development}
Jing, J., Sun, H., Kim, J.~A., Herlopian, A., Karakis, I., Ng, M., Halford,
  J.~J., Maus, D., Chan, F., Dolatshahi, M., et~al. (2020).
\newblock Development of expert-level automated detection of epileptiform
  discharges during electroencephalogram interpretation.
\newblock {\em JAMA neurology}, 77(1):103--108.

\bibitem[Katharopoulos et~al., 2020]{katharopoulos-et-al-2020}
Katharopoulos, A., Vyas, A., Pappas, N., and Fleuret, F. (2020).
\newblock Transformers are rnns: Fast autoregressive transformers with linear
  attention.
\newblock In {\em Proceedings of the International Conference on Machine
  Learning (ICML)}.

\bibitem[Kim et~al., 2020]{kim2020study}
Kim, M.-G., Ko, H., and Pan, S.~B. (2020).
\newblock A study on user recognition using 2d ecg based on ensemble of deep
  convolutional neural networks.
\newblock {\em Journal of Ambient Intelligence and Humanized Computing},
  11:1859--1867.

\bibitem[Klem et~al., 1999]{Klem1999TheTE}
Klem, G.~H., L{\"u}ders, H., Jasper, H.~H., and Elger, C.~E. (1999).
\newblock The ten-twenty electrode system of the international federation. the
  international federation of clinical neurophysiology.
\newblock {\em Electroencephalography and clinical neurophysiology.
  Supplement}, 52:3--6.

\bibitem[Kostas et~al., 2021]{kostas2021bendr}
Kostas, D., Aroca-Ouellette, S., and Rudzicz, F. (2021).
\newblock Bendr: using transformers and a contrastive self-supervised learning
  task to learn from massive amounts of eeg data.
\newblock {\em Frontiers in Human Neuroscience}, 15:653659.

\bibitem[Lawhern et~al., 2018]{lawhern2018eegnet}
Lawhern, V.~J., Solon, A.~J., Waytowich, N.~R., Gordon, S.~M., Hung, C.~P., and
  Lance, B.~J. (2018).
\newblock Eegnet: a compact convolutional neural network for eeg-based
  brain--computer interfaces.
\newblock {\em Journal of neural engineering}, 15(5):056013.

\bibitem[Li et~al., 2022]{li2022motor}
Li, H., Ding, M., Zhang, R., and Xiu, C. (2022).
\newblock Motor imagery eeg classification algorithm based on cnn-lstm feature
  fusion network.
\newblock {\em Biomedical signal processing and control}, 72:103342.

\bibitem[Lin et~al., 2017]{lin2017focal}
Lin, T.-Y., Goyal, P., Girshick, R., He, K., and Doll{\'a}r, P. (2017).
\newblock Focal loss for dense object detection.
\newblock In {\em Proceedings of the IEEE international conference on computer
  vision}, pages 2980--2988.

\bibitem[Liu et~al., 2021]{liu2021transformers}
Liu, J., Zhang, L., Wu, H., and Zhao, H. (2021).
\newblock Transformers for eeg emotion recognition.
\newblock {\em arXiv preprint arXiv:2110.06553}.

\bibitem[Liu et~al., 2019]{liu2019roberta}
Liu, Y., Ott, M., Goyal, N., Du, J., Joshi, M., Chen, D., Levy, O., Lewis, M.,
  Zettlemoyer, L., and Stoyanov, V. (2019).
\newblock Roberta: A robustly optimized bert pretraining approach.
\newblock {\em arXiv preprint arXiv:1907.11692}.

\bibitem[Lopez et~al., 2015]{lopez2015automated}
Lopez, S., Suarez, G., Jungreis, D., Obeid, I., and Picone, J. (2015).
\newblock Automated identification of abnormal adult eegs.
\newblock In {\em 2015 IEEE Signal Processing in Medicine and Biology Symposium
  (SPMB)}, pages 1--5. IEEE.

\bibitem[Nagabushanam et~al., 2020]{nagabushanam2020artifact}
Nagabushanam, P., George, S.~T., Davu, P., Bincy, P., Naidu, M., and Radha, S.
  (2020).
\newblock Artifact removal using elliptic filter and classification using
  1d-cnn for eeg signals.
\newblock In {\em 2020 6th International Conference on Advanced Computing and
  Communication Systems (ICACCS)}, pages 551--556. IEEE.

\bibitem[Nyquist, 1928]{nyquist1928certain}
Nyquist, H. (1928).
\newblock Certain topics in telegraph transmission theory.
\newblock {\em Transactions of the American Institute of Electrical Engineers},
  47(2):617--644.

\bibitem[OpenAI, 2023]{openai2023gpt4}
OpenAI (2023).
\newblock Gpt-4 technical report.

\bibitem[Parvaneh et~al., 2019]{parvaneh2019cardiac}
Parvaneh, S., Rubin, J., Babaeizadeh, S., and Xu-Wilson, M. (2019).
\newblock Cardiac arrhythmia detection using deep learning: A review.
\newblock {\em Journal of electrocardiology}, 57:S70--S74.

\bibitem[Peh et~al., 2022]{peh2022transformer}
Peh, W.~Y., Yao, Y., and Dauwels, J. (2022).
\newblock Transformer convolutional neural networks for automated artifact
  detection in scalp eeg.
\newblock In {\em 2022 44th Annual International Conference of the IEEE
  Engineering in Medicine \& Biology Society (EMBC)}, pages 3599--3602. IEEE.

\bibitem[Phan and Mikkelsen, 2022]{phan2022automatic}
Phan, H. and Mikkelsen, K. (2022).
\newblock Automatic sleep staging of eeg signals: recent development,
  challenges, and future directions.
\newblock {\em Physiological Measurement}.

\bibitem[Quan et~al., 1997]{quan1997sleep}
Quan, S.~F., Howard, B.~V., Iber, C., Kiley, J.~P., Nieto, F.~J., O'Connor,
  G.~T., Rapoport, D.~M., Redline, S., Robbins, J., Samet, J.~M., et~al.
  (1997).
\newblock The sleep heart health study: design, rationale, and methods.
\newblock {\em Sleep}, 20(12):1077--1085.

\bibitem[Sakhavi et~al., 2018]{sakhavi2018c2cm}
Sakhavi, S., Guan, C., and Yan, S. (2018).
\newblock Learning temporal information for brain-computer interface using
  convolutional neural networks.
\newblock {\em IEEE transactions on neural networks and learning systems},
  29(11):5619--5629.

\bibitem[Schirrmeister et~al., 2017]{schirrmeister2017deep}
Schirrmeister, R.~T., Springenberg, J.~T., Fiederer, L. D.~J., Glasstetter, M.,
  Eggensperger, K., Tangermann, M., Hutter, F., Burgard, W., and Ball, T.
  (2017).
\newblock Deep learning with convolutional neural networks for eeg decoding and
  visualization.
\newblock {\em Human brain mapping}, 38(11):5391--5420.

\bibitem[Shannon, 1949]{shannon1949communication}
Shannon, C.~E. (1949).
\newblock Communication in the presence of noise.
\newblock {\em Proceedings of the IRE}, 37(1):10--21.

\bibitem[Shoeb, 2009]{shoeb2009application}
Shoeb, A.~H. (2009).
\newblock {\em Application of machine learning to epileptic seizure onset
  detection and treatment}.
\newblock PhD thesis, Massachusetts Institute of Technology.

\bibitem[Song et~al., 2021]{song2021transformer}
Song, Y., Jia, X., Yang, L., and Xie, L. (2021).
\newblock Transformer-based spatial-temporal feature learning for eeg decoding.
\newblock {\em arXiv preprint arXiv:2106.11170}.

\bibitem[Srivastava et~al., 2014]{srivastava2014dropout}
Srivastava, N., Hinton, G., Krizhevsky, A., Sutskever, I., and Salakhutdinov,
  R. (2014).
\newblock Dropout: a simple way to prevent neural networks from overfitting.
\newblock {\em The journal of machine learning research}, 15(1):1929--1958.

\bibitem[Suhaimi et~al., 2020]{suhaimi2020eeg}
Suhaimi, N.~S., Mountstephens, J., Teo, J., et~al. (2020).
\newblock Eeg-based emotion recognition: A state-of-the-art review of current
  trends and opportunities.
\newblock {\em Computational intelligence and neuroscience}, 2020.

\bibitem[Vaswani et~al., 2017]{vaswani2017attention}
Vaswani, A., Shazeer, N., Parmar, N., Uszkoreit, J., Jones, L., Gomez, A.~N.,
  Kaiser, {\L}., and Polosukhin, I. (2017).
\newblock Attention is all you need.
\newblock {\em Advances in neural information processing systems}, 30.

\bibitem[Venkatachalam et~al., 2020]{venkatachalam2020novel}
Venkatachalam, K., Devipriya, A., Maniraj, J., Sivaram, M., Ambikapathy, A.,
  and Iraj, S.~A. (2020).
\newblock A novel method of motor imagery classification using eeg signal.
\newblock {\em Artificial intelligence in medicine}, 103:101787.

\bibitem[Wagner et~al., 2020]{wagner2020ptb}
Wagner, P., Strodthoff, N., Bousseljot, R.-D., Kreiseler, D., Lunze, F.~I.,
  Samek, W., and Schaeffter, T. (2020).
\newblock Ptb-xl, a large publicly available electrocardiography dataset.
\newblock {\em Scientific data}, 7(1):154.

\bibitem[Wang et~al., 2020]{wang2020linformer}
Wang, S., Li, B.~Z., Khabsa, M., Fang, H., and Ma, H. (2020).
\newblock Linformer: Self-attention with linear complexity.
\newblock {\em arXiv preprint arXiv:2006.04768}.

\bibitem[Yang et~al., 2022a]{yang2022atd}
Yang, C., Qian, C., Singh, N., Xiao, C.~D., Westover, M., Solomonik, E., and
  Sun, J. (2022a).
\newblock Atd: Augmenting cp tensor decomposition by self supervision.
\newblock {\em Advances in Neural Information Processing Systems},
  35:32039--32052.

\bibitem[Yang et~al., 2023]{yangmanydg}
Yang, C., Westover, M.~B., and Sun, J. (2023).
\newblock Manydg: Many-domain generalization for healthcare applications.
\newblock In {\em The Eleventh International Conference on Learning
  Representations}.

\bibitem[Yang et~al., 2022b]{yangpyhealth}
Yang, C., Wu, Z., Jiang, P., Lin, Z., and Sun, J. (2022b).
\newblock Pyhealth: A deep learning toolkit for healthcare predictive modeling,
  09 2022.
\newblock {\em URL https://github. com/sunlabuiuc/PyHealth}.

\bibitem[Yang et~al., 2021]{yang2021self}
Yang, C., Xiao, D., Westover, M.~B., and Sun, J. (2021).
\newblock Self-supervised eeg representation learning for automatic sleep
  staging.
\newblock {\em arXiv preprint arXiv:2110.15278}.

\bibitem[Zhang et~al., 2018]{zhang2018national}
Zhang, G.-Q., Cui, L., Mueller, R., Tao, S., Kim, M., Rueschman, M., Mariani,
  S., Mobley, D., and Redline, S. (2018).
\newblock The national sleep research resource: towards a sleep data commons.
\newblock {\em Journal of the American Medical Informatics Association},
  25(10):1351--1358.

\bibitem[Zhang et~al., 2019]{zhang2019cnnlstm}
Zhang, R., Zong, Q., Dou, L., and Zhao, X. (2019).
\newblock A novel hybrid deep learning scheme for four-class motor imagery
  classification.
\newblock {\em Journal of neural engineering}, 16(6):066004.

\bibitem[Zhang et~al., 2022]{zhang2022self}
Zhang, X., Zhao, Z., Tsiligkaridis, T., and Zitnik, M. (2022).
\newblock Self-supervised contrastive pre-training for time series via
  time-frequency consistency.
\newblock {\em arXiv preprint arXiv:2206.08496}.

\bibitem[Zhang et~al., 2020]{zhang2020investigation}
Zhang, Y., Chen, J., Tan, J.~H., Chen, Y., Chen, Y., Li, D., Yang, L., Su, J.,
  Huang, X., and Che, W. (2020).
\newblock An investigation of deep learning models for eeg-based emotion
  recognition.
\newblock {\em Frontiers in Neuroscience}, 14:622759.

\end{thebibliography}

\appendix

\section{Details of Datasets and Experimental Settings} 
\subsection{More for Datasets and Processings}\label{sec:datasets-details}
We provide more descriptions on each dataset in this section. 

{\bf For EEG datasets.} First, the 16 montages (in 10-20 international system) are "FP1-F7", "F7-T7", "T7-P7", "P7-O1", "FP2-F8", "F8-T8", "T8-P8", "P8-O2", "FP1-F3", "F3-C3", "C3-P3", "P3-O1", "FP2-F4", "F4-C4", "C4-P4", "P4-O2".
\begin{itemize}
    \item Sleep Heart Health Study ({\bf SHHS}) \citep{zhang2018national,quan1997sleep} is a multi-center cohort study from
the National Heart Lung \& Blood Institute assembled to study sleep-disordered breathing, which contains 5,445 recordings. The data is accessible upon request in their website \footnote{https://sleepdata.org/datasets/shhs}. Each recording has 14 Polysomnography (PSG) channels, and the recording frequency is 125.0 Hz. We use the C3/A2 and C4/A1 EEG channels. The dataset is released with sleep annotations. We use the existing codes \footnote{https://github.com/ycq091044/ContraWR/tree/main/preprocess} and split each recordings into 30-second samples. In this study, we use SHHS samples for unsupervised pre-training without it original labels.
\item {\bf PREST} is a private dataset recorded in hospital sleep lab, primarily for seizure and abnormal EEG detection purpose (such as spikes). The local IRB waived the requirement for informed consent for this retrospective analysis of EEG data. We follow the clinician's instructions and split each recordings into 10 seconds without labels. In the experiment, we use it for EEG model pre-training.
\item The {\bf CHB-MIT} database \footnote{https://physionet.org/content/chbmit/1.0.0/} \citep{shoeb2009application} is publicly available, which is collected at the Children’s Hospital Boston, consists of EEG recordings from pediatric subjects with intractable seizures. The dataset is under Open Data Commons Attribution License v1.0 \footnote{https://physionet.org/content/chbmit/view-license/1.0.0/} and is used to predict whether the EEG recordings contain seizure signals. Each recording initially contains 23 bipolar channels and we select the 16 standard montages in the experiments. We utilize the existing preprocessing \footnote{https://github.com/bernia/chb-mit-scalp} and follow the typical practices to further split each recordings into 10-second non-overlapping samples by default. Since the dataset is highly imbalanced, we use 5 seconds as overlaps to split the seizure regions (which could potentially double the positive samples). After processing, the positive ratio in the training set is around 0.6\%.
\item {\bf IIIC Seizure} is requested from \cite{jing2018rapid,ge2021deep,jing2023development}, and we follow the license and usage statements in \cite{jing2023development}. The samples follow 16 montages and span 10-second signals at 200Hz. This dataset is used for predicting one of the six classes: lateralized periodic discharges (LPD), generalized periodic discharges (GPD), lateralized rhythmic delta activity (LRDA), generalized rhythmic delta activity (GRDA), Seizure types, and Other.
\item TUH Abnormal EEG Corpus ({\bf TUAB}) \citep{lopez2015automated} and TUH EEG Events ({\bf TUEV}) \citep{harati2015improved} is accessible upon request at Temple University Electroencephalography (EEG) Resources \footnote{https://isip.piconepress.com/projects/tuh\_eeg/html/downloads.shtml}. We process both datasets to follow the 16 EEG montages.
\end{itemize}
{\bf For ECG datasets.} We use the Cardiology collection to pre-train the ECG models and apply it on downstream supervisd PTB-XL task.
\begin{itemize}
\item The {\bf Cardiology} collection \citep{alday2020classification} is publicly available at physionet \footnote{https://physionet.org/content/challenge-2020/1.0.2/}, which was used in the PhysioNet/Computing in Cardiology Challenge 2020. This collection is under Creative Commons Attribution 4.0 International Public License \footnote{https://physionet.org/content/challenge-2020/view-license/1.0.2/}. In this study, we use five sets from the training portion of the collection (It has in total six sets. Another one overlaps with the PTB-XL dataset, and thus we drop it in the pre-training), which contains recordings from CPSC2018 (6,877 recordings), CPSC2018Extra (China 12-Lead ECG Challenge Database – unused CPSC 2018 data, 3,453 recordings), St Petersburg Incart (12-lead Arrhythmia Database, 74 recordings), ptb (Diagnostic ECG Database, 516 recordings), Georgia (12-Lead ECG Challenge Database, 10,344 recordings). For preprocessing, we extract 10-second samples from each recording with 0.5s as the overlapping window. All the samples are merged together as an unsupervised pre-training ECG corpus of nearly 0.5 million samples. We pre-train a Pre-trained \method (Cardiology-12) on all the channels and a Pre-trained \method (Cardiology-6) on the first 6-channels of all samples. The sample sizes are different from the below PTB-XL dataset.
\item Physikalisch-Technische Bundesanstalt ({\bf PTB-XL})  \footnote{https://physionet.org/content/ptb-xl/1.0.1/} \citep{wagner2020ptb} is a publicly available large dataset of 12-lead ECGs from 18885 patients. It is under the Creative Commons Attribution 4.0 International Public License \footnote{https://physionet.org/content/ptb-xl/view-license/1.0.1/}. The raw waveform data was annotated by up to two cardiologists, who assigned potentially multiple ECG statements to each record up to 27 diagnoses: 1:1st degree AV block, 2:Atrial fibrillation, 3:Atrial flutter, 4:Bradycardia, 5:Complete right bundle branch block, 6:Incomplete right bundle branch block, 7:Left anterior fascicular block, 8:Left axis deviation, 9:Left bundle branch block, 10:Low QRS voltages, 11:Nonspecific intraventricular conduction disorder, 12:Pacing rhythm, 13:Premature atrial contraction, 14:Premature ventricular contractions, 15:Prolonged PR interval, 16:Prolonged QT interval, 17:Q wave abnormal, 18:Right axis deviation, 19:Right bundle branch block, 20:Sinus arrhythmia, 21:Sinus bradycardia, 22:Sinus rhythm, 23:Sinus tachycardia, 24:Supraventricular premature beats, 25:T wave abnormal, 26:T wave inversion, 27:Ventricular premature beats. We following clinical knowledges and further groups them into six broader categories: Arrhythmias, Bundle branch blocks and fascicular blocks, Axis deviations, Conduction delays, Wave abnormalities, Miscellaneous. Each recordings can be associated to multiple categories. In this paper, we conduct the "Arrhythmias" phenotyping prediction task. If the recordings have at least one diagnosis belonging to the Arrhythmias group, then we label them as positive, otherwise as negative.
\end{itemize}
{\bf For human activity sensory data.} Human activity recognition ({\bf HAR}) dataset \footnote{https://archive.ics.uci.edu/ml/datasets/human+activity+recognition+using+smartphones} \citep{anguita2013public} is publicly available at UCI machine learning repository. The data is collected from smartphone accelerometer and gyroscope data with 3D coordinates to detect six actions: walking, walking upstairs, walking downstairs, sitting, standing, laying. The samples are already splitted and provided in the original datasets.
\subsection{More for Experimental Settings} \label{sec:experiment-details}
For model implementation, the SPaRCNet code is requested from the authors \citep{jing2023development}, the ContraWR code is downloaded and modified upon the github \footnote{https://github.com/ycq091044/ContraWR}, CNN-Transformer is easily implemented following the Fig. 3 of the original paper \citep{peh2022transformer}, FFCL \citep{li2022motor} combines a CNN model and a LSTM model for learning separete representations and then merges them before the final prediction layer, the implementation of ST-Transformer refer to this repo \footnote{https://github.com/eeyhsong/EEG-Transformer}. The linear-complexity attention module is referred to this repo \footnote{https://github.com/lucidrains/linear-attention-transformer} in our \method implementation.


For all EEG tasks, we resample the datasets into 200Hz. The ECG tasks use 500Hz, and the HAR tasks use 50Hz by default. For each specific tasks, we have to adjust the baseline model architectures (e.g, number of layers, input channel sizes, etc) accordingly since the input data have various formats. While for our \method, we only adjust the fft size based on their sampling rate (200 points for EEG, 1000 points for ECG, 100 points for HAR) and use 100 points, 200 points, and 10 points as the hop length (i.e., overlaps) in three signal types. These configurations are chosen by testing several other combinations based on the validation performance. For our \method model, we use 8 as the number of head, 4 as the number of transformer layers, and $T=2$ as the temperature in unsupervised pre-training by default. We use the Adam optimizer with learning rate $1\times 10^{-3}$ and $1\times 10^{-5}$ as the coefficient for L2 regularization by default. We use the pytorch lightning framework (with 100 as the max epoch) to handle the training, validation, and test pipeline by setting AUROC as the monitoring metirc for binary classification and Coken's Kappa as the monitoring metric for multi-class classification. More details can refer to our Supplementary codes. Below, we provide the definition of each metric used in the paper, and we use {\em pyhealth.metrics} \footnote{https://pyhealth.readthedocs.io/en/latest/api/metrics.html} \cite{yangpyhealth} module for the implementation.

{\bf Balanced Accuracy} is defined as the average of recall obtained on each class. It is used for both binary classification and multi-class classification.

{\bf AUC-PR} is the area under the precision recall (PR) curve for binary classification task.

{\bf AUROC} is the area under the ROC curve, summarizing the ROC curve into an single number that describes the performance of a model for multiple thresholds at the same time. It is used for binary classification.

{\bf Coken’s Kappa} is a statistic that measures inter-annotator agreement, which is usually used for imbalanced multi-class classification task. The calculation can refer to sklearn metrics \footnote{https://scikit-learn.org/stable/modules/generated/sklearn.metrics.cohen\_kappa\_score.html}.

{\bf Weighted F1} is used for multi-class classification in this paper, which is a weighted average of individual F1-scores from each class, with each score weighted by the number of samples in the corresponding class.

\section{Additional Results}
This section provides additional experimental results to support claims in the main paper.
\subsection{Additional Experiments on TUEV and TUAB}\label{sec:eeg-task2}
We have provided the supervised learning results on EEG dataset IIIC Seizure and CHB-MIT in the main text. For completeness, we provide similar comparison results on TUAB and TUEV below in Table~\ref{tb:eeg-task2}~\ref{tb:tuev}, which show a similar trend that our \method shows better performance against baseline models, and the pre-trained \method models can bring significant improvements on two downstream tasks, especially on TUEV. For TUEV, we also append the results of all different pre-trained models (e.g., train from scratch, supervised training, unsupervised training, etc) in the end in Table~\ref{tb:tuev}.
\begin{table}[h!]
\centering
\caption{Additional Supervised Learning Results on TUAB}
\resizebox{0.75\textwidth}{!}{\begin{tabular}{l|ccc} 
\toprule 
  \multirow{2}{*}{\bf Models} & \multicolumn{3}{c}{\bf TUAB (abnormal detection)} \\
  \cmidrule{2-4}
  & Balanced Acc. & AUC-PR & AUROC \\
  \midrule
SPaRCNet        & 0.7896 $\pm$ 0.0018 & 0.8414 $\pm$ 0.0018 & 0.8676 $\pm$ 0.0012\\
ContraWR        & 0.7746 $\pm$ 0.0041 & 0.8421 $\pm$ 0.0104 & 0.8456 $\pm$ 0.0074 \\
CNN-Transformer & 0.7777 $\pm$ 0.0022 & 0.8433 $\pm$ 0.0039 & 0.8461 $\pm$ 0.0013 \\
FFCL            & 0.7848 $\pm$ 0.0038 & 0.8448 $\pm$ 0.0065 & 0.8569 $\pm$ 0.0051 \\
ST-Transformer  & 0.7966 $\pm$ 0.0023 & 0.8521 $\pm$ 0.0026 & {\bf 0.8707 $\pm$ 0.0019}\\
(Vanilla) BIOT  & {\bf 0.7925 $\pm$ 0.0035} & {\bf 0.8707 $\pm$ 0.0087} & 0.8691 $\pm$ 0.0033 \\
\midrule
Pre-trained BIOT (PREST) & 0.7907 $\pm$ 0.0050 & {0.8752 $\pm$ 0.0051} & 0.8730 $\pm$ 0.0021  \\
Pre-trained BIOT (PREST+SHHS) & \fbox{0.8019 $\pm$ 0.0021} & {0.8749 $\pm$ 0.0054} & {0.8739 $\pm$ 0.0019} \\
Pre-trained BIOT (6 EEG datasets) & 0.7959 $\pm$ 0.0057 & \fbox{0.8792 $\pm$ 0.0023} & \fbox{0.8815 $\pm$ 0.0043} \\
\bottomrule
\multicolumn{4}{l}{* {\bf Bold} for the best model (trained from scratch) and \fbox{box} for the best pre-trained models.}
\end{tabular}}
\label{tb:eeg-task2} 
\end{table}

\begin{table}[h!]
\centering
\caption{Additional Supervised Learning Results on TUEV (All-in-one-table comparison)}
\resizebox{1.02\textwidth}{!}{\begin{tabular}{l|ccc} 
\toprule 
  \multirow{2}{*}{\bf Models} & \multicolumn{3}{c}{\bf TUEV (event type classification)}\\
  \cmidrule{2-4}
  & Balanced Acc. & Coken’s Kappa & Weighted F1\\
  \midrule
  \multicolumn{4}{l}{\bf (Training from scratch in Section~\ref{sec:supervised-learning}} \\
  \midrule
SPaRCNet        & 0.4161 $\pm$ 0.0262 & 0.4233 $\pm$ 0.0181 & 0.7024 $\pm$ 0.0104\\
ContraWR        & 0.4384 $\pm$ 0.0349 & 0.3912 $\pm$ 0.0237 & 0.6893 $\pm$ 0.0136\\
CNN-Transformer & 0.4087 $\pm$ 0.0161 & 0.3815 $\pm$ 0.0134 & 0.6854 $\pm$ 0.0293 \\
FFCL            & 0.3979 $\pm$ 0.0104 & 0.3732 $\pm$ 0.0188 & 0.6783 $\pm$ 0.0120 \\
ST-Transformer  & 0.3984 $\pm$ 0.0228 & 0.3765 $\pm$ 0.0306 & 0.6823 $\pm$ 0.0190\\
(Vanilla) BIOT  & {0.4682 $\pm$ 0.0125} & {0.4482 $\pm$ 0.0285} & {0.7085 $\pm$ 0.0184} \\
\midrule
\multicolumn{4}{l}{\bf (Unsupervised pre-trained models in Section~\ref{sec:unsupervised-pre-train}):} \\
\midrule
Pre-trained BIOT (PREST) & {0.5207 $\pm$ 0.0285} & 0.4932 $\pm$ 0.0301 & 0.7381 $\pm$ 0.0169 \\
Pre-trained BIOT (PREST+SHHS) & 0.5149 $\pm$ 0.0292 & 0.4841 $\pm$ 0.0309 & 0.7322 $\pm$ 0.0196 \\
\midrule
\multicolumn{4}{l}{\bf (Supervised pre-trained models in Section~\ref{sec:supervised-pre-train}):} \\
\midrule
Pre-trained BIOT (pre-trained on CHB-MIT with 8 channels and 10s) & 0.4123 $\pm$ 0.0087 & 0.4285 $\pm$ 0.0065 & 0.6989 $\pm$ 0.0015 \\
Pre-trained BIOT (pre-trained on CHB-MIT with 16 channels and 5s) & 0.4218 $\pm$ 0.0117 & 0.4427 $\pm$ 0.0093 & 0.7147 $\pm$ 0.0058 \\
Pre-trained BIOT (pre-trained on CHB-MIT with 16 channels and 10s) & 0.4344 $\pm$ 0.0065 & 0.4719 $\pm$ 0.0231 & 0.7280 $\pm$ 0.0126 \\
Pre-trained BIOT (pre-trained on IIIC seizure with 8 channels and 10s) & 0.4956 $\pm$ 0.0552 & 0.4719 $\pm$ 0.0475 & 0.7214 $\pm$ 0.0220 \\
Pre-trained BIOT (pre-trained on IIIC seizure with 16 channels and 5s) & 0.4894 $\pm$ 0.0189 & 0.4881 $\pm$ 0.0045 & 0.7348 $\pm$ 0.0056 \\
Pre-trained BIOT (pre-trained on IIIC seizure with 16 channels and 10s) & 0.4935 $\pm$ 0.0288 & 0.5316 $\pm$ 0.0176 & 0.7555 $\pm$ 0.0111 \\
Pre-trained BIOT (pre-trained on TUAB with 8 channels and 10s) & 0.4980 $\pm$ 0.0384 & 0.4487 $\pm$ 0.0535 & 0.7044 $\pm$ 0.0365 \\
Pre-trained BIOT (pre-trained on TUAB with 16 channels and 5s) & 0.4954 $\pm$ 0.0305 & 0.5053 $\pm$ 0.0079 & 0.7447 $\pm$ 0.0049 \\
Pre-trained BIOT (pre-trained on TUAB with 16 channels and 10s) & 0.5256 $\pm$ 0.0348 & 0.5187 $\pm$ 0.0160 & 0.7504 $\pm$ 0.0102 \\
\midrule
\multicolumn{4}{l}{\bf (Supervised + unsupervised pre-trained model in Section~\ref{sec:ultimate}):} \\
\midrule
Pre-trained BIOT (ultimate) & {0.5281 $\pm$ 0.0225} & {0.5273 $\pm$ 0.0249} & {0.7492 $\pm$ 0.0082}\\
\bottomrule
\end{tabular}}
\label{tb:tuev} 
\end{table}

\subsection{Additional Experiments on CHB-MIT}\label{sec:chb-mit}
This section performs a similar experiment on CHB-MIT, similar to Section~\ref{sec:supervised-learning}. We pre-train on  the training set of IIIC Seizure (which has 16 channels and 10s duration), TUAB (which has 16 channels and 10s duration), TUEV (which has 16 channels and 5s duration) and fine-tunes on CHB-MIT (which has 16 channels and 10s duration). All datasets use 200Hz sampling rate. We design five sets of configurations for the pre-trained datasets: {\bf Format (i)} uses the first 8 channels and 10s duration; {\bf Format (ii)} uses the full 16 channels but only the first 5s recording; {\bf Format (iii)} uses full 16 channels and full 10s recording; {\bf Format (iv)} uses 8 channels and 5s recording, and {\bf Format (v)} uses full 16 channels and 2.5s recording. The last two are only for the TUEV dataset. During fine-tuning, we then remove the prediction layers from these pre-trained model and add a new prediction layer to fit the CHB-MIT dataset. 

\begin{figure}[th!]
    \includegraphics[width=0.98\textwidth]{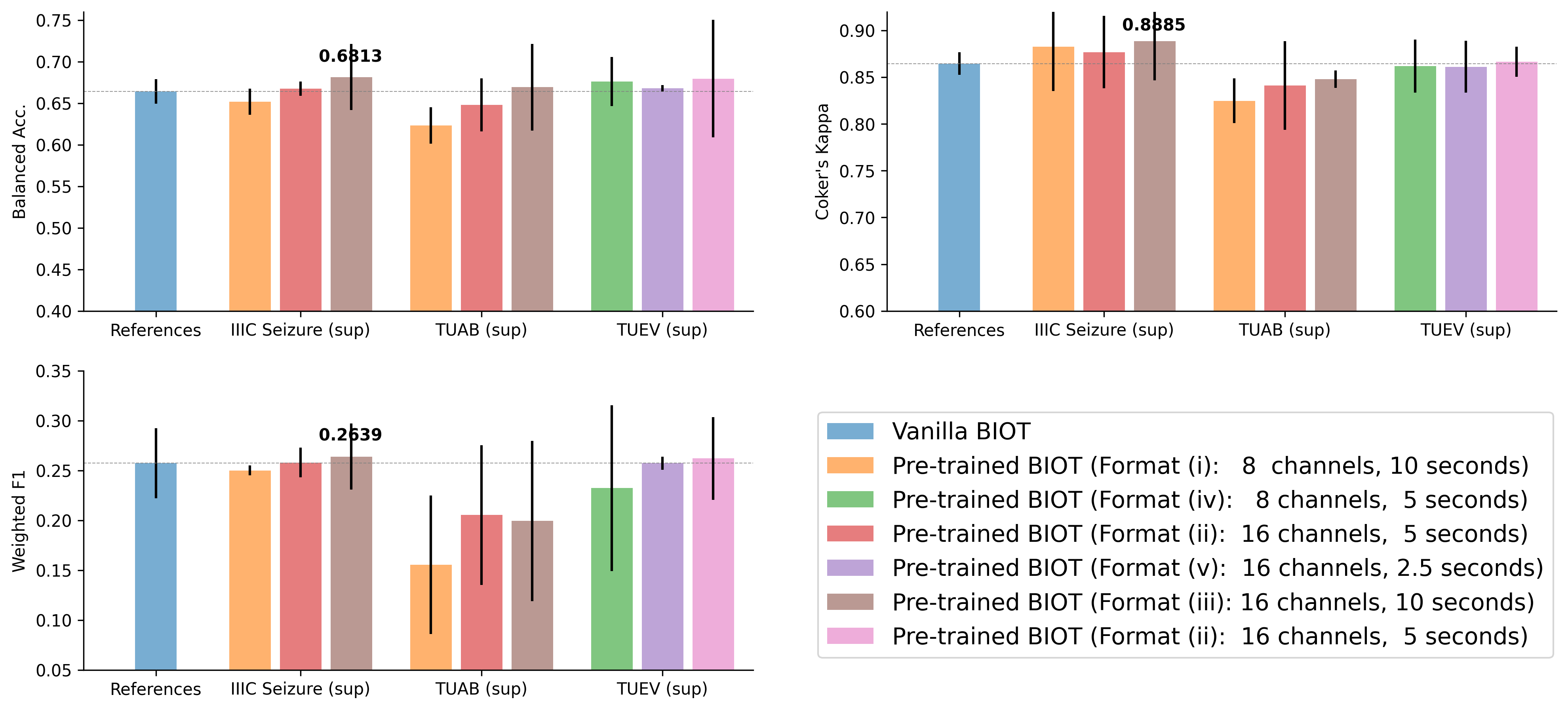}
    \vspace{-1mm}
    \caption{Fine-tuned on CHB-MIT from different supervised pre-trained models. IIIC Seizure and TUAV datasets follow Format (i)(ii)(iii), while TUEV follows the Format (iv)(v)(ii).}
    \label{fig:transfer2}
\end{figure}

The results are reported in Figure~\ref{fig:transfer2}, which shows that the supervised pre-training on both IIIC seizure and TUEV can help improve the downstream performance on CHB-MIT task compared to training from scratch. The reason is that IIIC Seizure is on multiple seizure type classification while CHB-MIT is on binary seizure or not classification, and the context of both tasks are fairly related. Although TUEV is not entirely on seizure related classification, some classes in TUEV are seizure subtypes (such as GPED, PLED), and thus its supervisd pre-trained models can also bring benefits for the CHB-MIT task.

\subsection{Ablation Studies on Hyperparameters} \label{sec:ablation-study}
This section provides ablation studies on three hyperparameters in data processing: target sampling rate, token duration, and the overlap size between two neighboring tokens. We use two EEG datasets as example: IIIC Seizure and TUAB. The default configuration in the main paper is {\bf (1) sampling}: 200Hz, {\bf (2) token length}: 1s, {\bf (3) overlaps}: 0.5s as reference. 
\subsubsection{Ablation Study on Target Sampling Rate $r$} \label{sec:target-sampling-rate}
In this experiments, we fix (2)(3) and conduct ablation study on the target sampling rate. The original IIIC Seizure data is at 200Hz and the TUAB data is at 256Hz. For IIIC Seizure, we vary the sampling rate to 26Hz, 50Hz, 100Hz, 150Hz, and 200Hz. For TUAB, we vary the sampling rate to 50Hz, 100Hz, 150Hz, 200Hz, 250Hz, and 300Hz. The evaluations are conducted under three different random seeds and the mean and standard deviation values are reported. 

For IIIC Seizure, we can observe that a higher sampling rate could give slightly better performance, especially on balanced acc. and coken's kappa. The reason is that higher sampling rate can preserve more detailed (high-frequency) biosignal information. The results on TUAB shows that the performances are similar on all sampling rates. We conjecture that different tasks might have diverse sensitivity to the the frequency bands. For example, the task on IIIC seizure is to classify different seizure types, which may need to capture minor clues from high-frequency waves (such as Gamma waves (30-100Hz)), while the TUAB dataset is for abnormal detection, and using brain waves under 50Hz might be enough for the task. In sum, the target sampling rate should be selected based on the predicting targets.
\begin{figure}[h!]
    \includegraphics[width=0.95\textwidth]{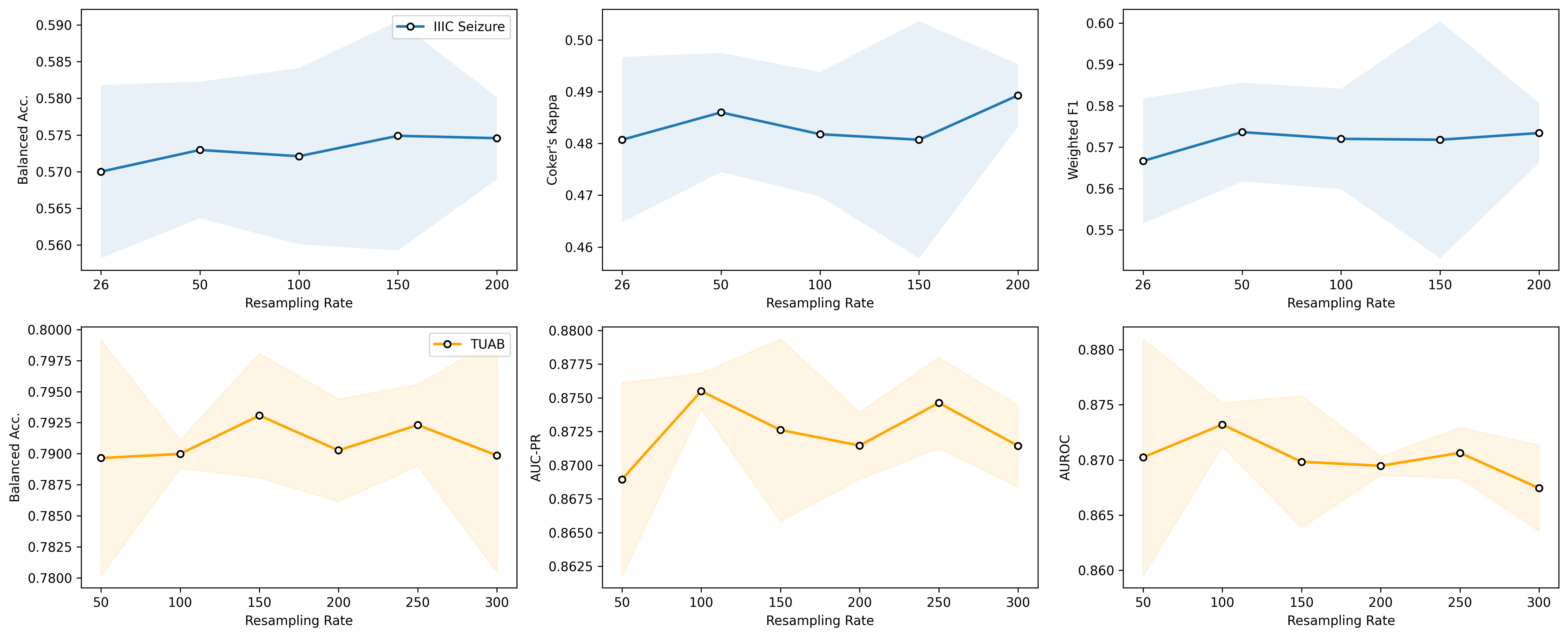}
    \caption{Ablation Study on Target Sampling Rate $r$}
    \label{fig:resampling-rate}
\end{figure}

\subsubsection{Ablation Study on Token Lengths $t$} \label{sec:token-length}
In this experiments, we fix (1)(3) and conduct ablation study on the token length. Both datasets have 10s as the entire sample length and 0.5s as the overlap lengths. For both of them, we vary the token lengths to 0.75s, 1s, 1.5s, 2s, 2.5s, 5s. The evaluations are conducted under three different random seeds and the mean and standard deviation values are reported. 

For each configuration, we also vary the fft size to match the token length, which means that 5s token length can extract more frequency information. However, we find that by increasing the token lengths, the model performance starts to decrease. Performances on IIIC Seizure starts to decrease after 1s while the performance on TUAB decreases after 2s. The reason could be that given the increaseing token lengths $t$, the total biosignal "sentence" length, which is $\frac{J~-t}{t-p}+1=\frac{10-t}{t-0.5}+1$, will decrease (here, $J$ is the channel biosignal duration, $t$ is the token length, $p$ is the overlapping length). For example, with $x=5$ as the token lengths, the final "sentence" length becomes $11$ while it is $19$ in the default configuration with $x=1s$. The performance drops is due to transformer models will be less beneficial in shorter "sentence"s.
\begin{figure}[h!]
    \includegraphics[width=0.95\textwidth]{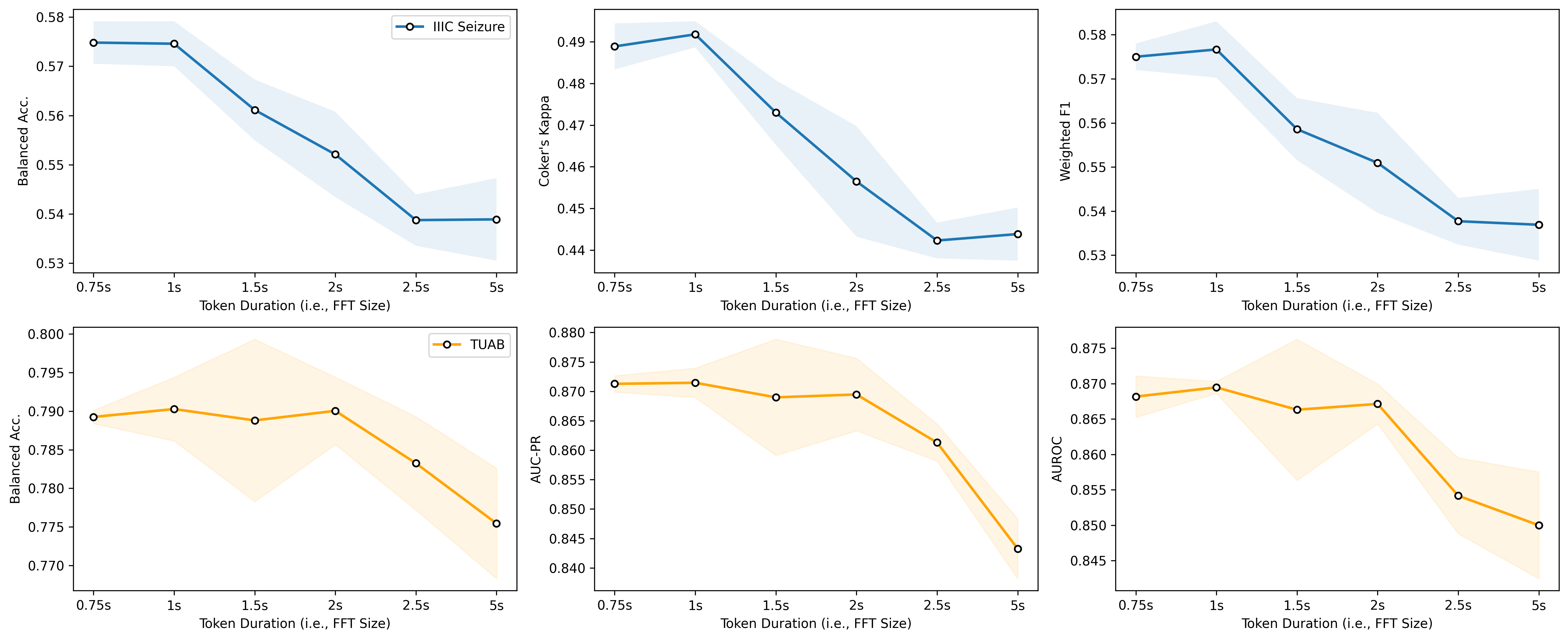}
    \caption{Ablation Study on Token Lengths $t$}
    \label{fig:token-length}
\end{figure}
\subsubsection{Ablation Study on Overlapping Lengths $p$} \label{sec:overlapping-length}
In this experiments, we fix (1)(2) and conduct ablation study on the overlap lengths. Both datasets have 10s as the entire sample length and 1s as the token lengths. For both of them, we vary the overlap lengths to 0.875s, 0.75s, 0.5s, 0.25s, 0s. The evaluations are conducted under three different random seeds and the mean and standard deviation values are reported. 

Based on the "sentence" length formula $\frac{J-t}{t-p}+1$, smaller overlap lengths will decrease the "sentence" length. On both datasets, we find that larger overlaps can brings slightly better results due to that the biosignal "sentence" becomes longer. Another reason is that with larger overlaps, neighboring tokens can capture more transitioning information and help the transformer model to better capture the temporal information.

\begin{figure}[h!]
    \includegraphics[width=0.95\textwidth]{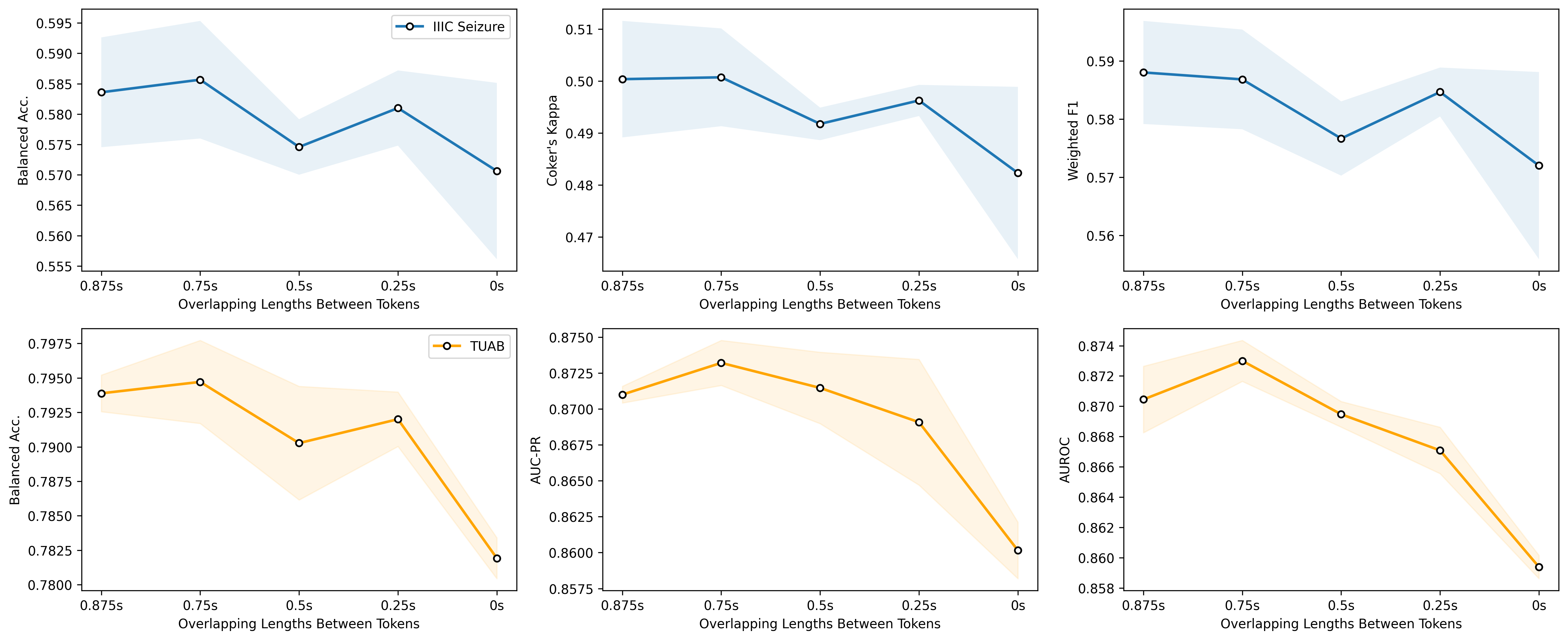}
    \caption{Ablation Study on Overlapping Lengths $p$ Between Tokens}
    \label{fig:hop-length}
\end{figure}

\end{document}